%% file: ms.tex
\documentclass{emulateapj}

\usepackage{graphicx}
\usepackage{amsmath}
\usepackage{enumitem}
\usepackage{longtable}

\newcommand{\ms}{\ifmmode{\rm m\thinspace s^{-1}}\else m\thinspace s$^{-1}$\fi}
\newcommand{\kms}{\ifmmode{\rm km\thinspace s^{-1}}\else km\thinspace s$^{-1}$\fi}
\newcommand{\kepler}{{\it Kepler\/}}
\newcommand{\Kepler}{{\it KEPLER\/}}
\newcommand\blender{{\tt BLENDER}}
\newcommand\vespa{{\tt VESPA}}

\shorttitle{Validation of small HZ planets}
\shortauthors{Torres et al.}

\begin{document}

\submitted{Accepted for publication in The Astronomical Journal}


\title{Validation of small \Kepler\ transiting planet candidates \\ in or
  near the habitable zone}

\author{
Guillermo Torres\altaffilmark{1},
Stephen R.\ Kane\altaffilmark{2}
Jason F.\ Rowe\altaffilmark{3},
Natalie M.\ Batalha\altaffilmark{4}
Christopher E.\ Henze\altaffilmark{4},
David R.\ Ciardi\altaffilmark{5},
Thomas Barclay\altaffilmark{6},
William J.\ Borucki\altaffilmark{4},
Lars A.\ Buchhave\altaffilmark{7},
Justin R.\ Crepp\altaffilmark{8},
Mark E.\ Everett\altaffilmark{9},
Elliott P.\ Horch\altaffilmark{10,11},
Andrew W.\ Howard\altaffilmark{12},
Steve B.\ Howell\altaffilmark{4},
Howard T.\ Isaacson\altaffilmark{13},
Jon M.\ Jenkins\altaffilmark{4},
David W.\ Latham\altaffilmark{1},
Erik A.\ Petigura\altaffilmark{12,14}, and
Elisa V.\ Quintana\altaffilmark{6}
}

\altaffiltext{1}{Harvard-Smithsonian Center for Astrophysics, 60 Garden
  Street, Cambridge, MA 02138, USA; gtorres@cfa.harvard.edu}

\altaffiltext{2}{Department of Earth Sciences, University of
  California, Riverside, CA 92521, USA}

\altaffiltext{3}{Department of Physics and Astronomy, Bishop's
  University, 2600 College St., Sherbrooke, QC, J1M 1Z7, Canada}

\altaffiltext{4}{NASA Ames Research Center, Moffett Field, CA 94035, USA}

\altaffiltext{5}{NASA Exoplanet Science Institute/Caltech, Pasadena, CA
  91125, USA}

\altaffiltext{6}{NASA Goddard Space Flight Center, Greenbelt, MD 20771,
  USA}

\altaffiltext{7}{Centre for Star and Planet Formation, Natural History
  Museum of Denmark \& Niels Bohr Institute, University of Copenhagen,
  DK-1350 Copenhagen K, Denmark}

\altaffiltext{8}{Department of Physics, University of Notre Dame, Notre
  Dame, IN 46556, USA}

\altaffiltext{9}{National Optical Astronomy Observatory, Tucson, AZ
  85719, USA}

\altaffiltext{10}{Department of Physics, Southern Connecticut State
  University, New Haven, CT 06515, USA}

\altaffiltext{11}{Adjunct Astronomer, Lowell Observatory, Flagstaff, AZ
  86001, USA}

\altaffiltext{12}{California Institute of Technology, Pasadena, CA
  91125, USA}

\altaffiltext{13}{Astronomy Department, University of California,
  Berkeley, CA 94720, USA}

\altaffiltext{14}{NASA Hubble Fellow}

\begin{abstract}

A main goal of NASA's \kepler\ Mission is to establish the frequency
of potentially habitable Earth-size planets ($\eta_{\earth}$).
Relatively few such candidates identified by the mission can be
confirmed to be rocky via dynamical measurement of their mass. Here
we report an effort to validate 18 of them statistically using
the \blender\ technique, by showing that the likelihood they are true
planets is far greater than that of a false positive. Our analysis
incorporates follow-up observations including high-resolution optical
and near-infrared spectroscopy, high-resolution imaging, and
information from the analysis of the flux centroids of the
\kepler\ observations themselves. While many of these candidates have
been previously validated by others, the confidence levels reported
typically ignore the possibility that the planet may transit a
different star than the target along the same line of sight. If that
were the case, a planet that appears small enough to be rocky may
actually be considerably larger and therefore less interesting from
the point of view of habitability. We take this into consideration
here, and are able to validate 15 of our candidates at a 99.73\%
(3$\sigma$) significance level or higher, and the other three at
slightly lower confidence. We characterize the GKM host stars using
available ground-based observations and provide updated parameters for
the planets, with sizes between 0.8 and 2.9\;$R_{\earth}$.
Seven of them (KOI-0438.02, 0463.01, 2418.01, 2626.01, 3282.01,
4036.01, and 5856.01) have a better than 50\% chance of being
smaller than 2\;$R_{\earth}$ and being in the habitable zone of their
host stars.

\end{abstract}


\section{Introduction}
\label{sec:introduction}

The occurrence rate of terrestrial-size planets within the habitable
zone (HZ) of their host stars, referred to as ``eta Earth'', or
$\eta_{\earth}$, is one of the fundamental quantities that the
exoplanet community is focusing their efforts on. The vast numbers of
transiting planet candidates from the \kepler\ Mission
\citep{Borucki:2016} are the primary source for these calculations,
and there have been many efforts to estimate the value of
$\eta_{\earth}$ from those data \citep[see the recent examples of][and
  references therein]{Dressing:2013, Kopparapu:2013a, Burke:2015,
  Dressing:2015, Mulders:2015}. A key aspect of determining the
reliability of these estimates is the confirmation of
\kepler\ candidates (\kepler\ Objects of Interest, or KOIs),
particularly for earlier-type stars for which the orbital periods for
the habitable zone become increasingly longer and the data more prone
to false positives \citep{Burke:2015, Coughlin:2016}.

Many of the \kepler\ stars that appear to host small planets in the HZ
are faint, or have other properties such as significant rotation or
chromospheric activity that make it very difficult to obtain the
high-precision radial-velocity measurements needed for a dynamical
confirmation of the planetary nature of the candidate. KOIs with long
orbital periods ($P$) are even more challenging as the radial-velocity
amplitudes fall off as $P^{-1/3}$, resulting in Doppler signals that
are often of the order of 1\;\ms\ or less, which is at the limit of
the detection capabilities of present instrumentation and techniques.

\cite{Kane:2016} have recently published a catalog of transiting HZ
candidates from \kepler\ based on the best available set of planetary
and stellar parameters available to them. Many of these candidates are
nominally smaller than 2\;$R_{\earth}$, and have relatively long
orbital periods up to several hundred days. These KOIs are therefore
of great interest in connection with efforts to establish
$\eta_{\earth}$, and yet most of them have remained unconfirmed for
one or more of the reasons mentioned above.

An alternative to dynamical confirmation is statistical validation, in
which the goal is to show that the likelihood of a false positive is
much smaller than that of a true planet. A number of the candidates
presented by \cite{Kane:2016} have been validated by others, and have
subsequently received official \kepler\ planet designations. However,
in most cases those validation studies have only been concerned with
demonstrating the presence of a planet associated with the target, but
not necessarily \emph{orbiting\/} it. In particular, in reporting a
confidence level for the validation they have usually not accounted
for the possibility that the planet may instead transit an unresolved
star near the target, either physically bound to it, or a chance
alignment \citep{Lissauer:2014, Morton:2016}. Such situations can in
fact be more common than the types of false positives normally
considered in these validation studies, by one to three orders of
magnitude \citep[see, e.g.,][]{Fressin:2013, Torres:2015}. If the
planet orbits a different star, the transit signal observed would not
reflect the true size of the planet. Instead, the true size could be
considerably larger \citep{Ciardi:2015, Furlan:2017}, possibly
implying an icy or gaseous composition rather than a rocky one. This
would make the planet less interesting from the standpoint of
habitability and $\eta_{\earth}$.

The motivation for the present work is thus to examine each of the
most promising \kepler\ HZ candidates more closely and revisit the
validations with attention to this issue, making use of follow-up
observations and other constraints not previously available. We use
these observations to also provide updated parameters for the
validated planets.

The paper is organized as follows. The target selection for this work
is explained in Section\;\ref{sec:sample}, followed by a description
of the \kepler\ photometry we use. Section\;\ref{sec:followup}
presents the follow-up observations for the targets, which includes
high-resolution imaging, an analysis of the motion of the flux
centroids, and high-resolution spectroscopy. Then in
Section\;\ref{sec:stellarproperties} we describe our analysis of the
spectroscopic material to determine the stellar properties of the host
stars (temperatures, metallicities, masses, radii, mean densities,
ages, etc.). The statistical validation procedure and results are
presented in Section\;\ref{sec:validation}, after which we report our
light-curve fits that yield the planetary parameters
(Section\;\ref{sec:fits}). The habitability of the planets is
discussed in Section\;\ref{sec:habitability}, and the last section
features our concluding remarks.

\section{Target selection}
\label{sec:sample}

The source of our target list is an early version of the catalog of
small HZ candidates published by \cite{Kane:2016}, which in turn is
the product of the efforts by the \kepler\ HZ Working Group to
evaluate the full set of candidates observed during the mission's
quarters 1 through 17 (Q1--Q17). This catalog describes the various
definitions used by different authors for the HZ, and chose to
consider both an ``optimistic'' (larger) HZ and a ``conservative''
(smaller) HZ, based on the assumptions of \cite{Kopparapu:2014}
informed by estimates of how long Venus and Mars may have been able to
retain liquid water on their surfaces. For reference, the inner and
outer boundaries adopted for the optimistic HZ correspond to
approximately 0.75 and 1.8\;au for a star like the Sun, and those of the
conservative HZ are located at about 0.99 and 1.7\;au, although these
limits vary depending on the exact temperature of the star because of
changes in the albedo of an Earth-like planet under the different
wavelengths of stellar irradiation.  The compilation of
\cite{Kane:2016} also separated the candidates according to their size
(planetary radius $R_p$), defining four categories as follows, with
some being subsets of others:
\begin{enumerate}[leftmargin=15pt,itemsep=-3.5pt]
\item Candidates in the conservative HZ with $R_p \le 2\:R_{\earth}$;
\item Candidates in the optimistic HZ with $R_p \le 2\:R_{\earth}$;
\item Candidates in the conservative HZ with any radius;
\item Candidates in the optimistic HZ with any radius.
\end{enumerate}

The present work began during the early stages of preparation of the
catalog of \cite{Kane:2016} with the selection of 19 candidates for
validation from Categories\;1 and 2, ranked of high interest based on
their small size and likelihood of being in the HZ. However, due in
part to subsequent improvements in the stellar parameters
(particularly the stellar radii and temperatures) that led to revised
planetary parameters, the KOIs considered for inclusion in the
\cite{Kane:2016} catalog evolved with time until its publication, and
as a result not all of the targets we initially selected for
validation ended up in the final version of the catalog. On the other
hand, out of concern that some of the signals might be spurious, we
had originally chosen to exclude candidates in the catalog with low or
marginal signal-to-noise ratios as represented by the Multiple Event
Statistic (MES) listed on NASA's Exoplanet Archive\footnote{\url
  http://exoplanetarchive.ipac.caltech.edu/}. The MES measures the
significance of the observed transits in the detrended, whitened light
curve \citep{Jenkins:2002}. We rejected KOIs with MES values lower
than about 10, based on the estimates from the \kepler\ data release
current at the time \citep[Q1--Q17 Data Release\;24, or
  DR24;][]{Coughlin:2016}. The most recent and final data release
\citep[DR25;][]{Thompson:2017} did not alter that selection except in
the case of KOI-7235.01, which was accepted by \cite{Kane:2016} but is
now considered to be a false alarm. We therefore dropped this
candidate from the target list, and retained the other 18. Six of
these are in Category\;1, three additional ones are in Category\;2,
two more are in Category\;3, and one is in Category\;4. The other six
of our original targets are not listed in any of the categories of the
catalog of \cite{Kane:2016}.

Table\;\ref{tab:targets} presents the set of targets we have kept for
this study, and includes the \kepler\ planet designation when the
candidate has been statistically validated previously or as a result
of this work, along with the MES and transit depth as listed on the
Exoplanet Archive, and other ancillary information.

\begin{deluxetable*}{llrccccc}
\tablewidth{0pc}
\tablecaption{Sample of KOIs in this Study.\label{tab:targets}}
\tablehead{
\colhead{} &
\colhead{} &
\colhead{} &
\colhead{$K\!p$} &
\colhead{$b$} &
\colhead{Period} &
\colhead{} &
\colhead{Depth}
\\
\colhead{Candidate} &
\colhead{Name} &
\colhead{KID} &
\colhead{(mag)} &
\colhead{(deg)} &
\colhead{(days)} &
\colhead{MES} &
\colhead{(ppm)}
}
\startdata
 KOI-0172.02  &   Kepler-69\,c  (1)    &      8692861  &  13.749  &    +11.99  &    242.47  &    18.0   &  \phn340  \\
 KOI-0438.02  &   Kepler-155\,c (2)    &     12302530  &  14.258  &    +17.45  & \phn52.66  &    30.6   &     1078  \\
 KOI-0463.01  &   Kepler-560\,b (3)    &      8845205  &  14.708  & \phn+7.76  & \phn18.48  &    78.0   &     2605  \\
 KOI-0812.03  &   Kepler-235\,e (2)    &      4139816  &  15.954  &    +14.48  & \phn46.18  &    18.0   &     1395  \\
 KOI-0854.01  &   Kepler-705\,b (3)    &      6435936  &  15.849  &    +13.13  & \phn56.06  &    19.3   &     1626  \\
 KOI-2418.01  &   Kepler-1229\,b (3)   &     10027247  &  15.474  &    +10.38  & \phn86.83  &    11.7   &  \phn751  \\
 KOI-2626.01  &   Kepler-1652\,b (4)   &     11768142  &  15.931  &    +13.57  & \phn38.10  &    14.6   &  \phn913  \\
 KOI-2650.01  &   Kepler-395\,c (2)    &      8890150  &  15.987  &    +11.96  & \phn34.99  &    10.1   &  \phn498  \\
 KOI-3010.01  &   Kepler-1410\,b (3)   &      3642335  &  15.757  &    +11.13  & \phn60.87  &    12.7   &  \phn714  \\
 KOI-3282.01  &   Kepler-1455\,b (3)   &     12066569  &  15.855  &    +13.99  & \phn49.28  &    14.7   &     1133  \\
 KOI-3497.01  &   Kepler-1512\,b (3)   &      8424002  &  13.393  &    +14.39  & \phn20.36  &    19.6   &  \phn346  \\
 KOI-4036.01  &   Kepler-1544\,b (3)   &     11415243  &  14.061  &    +11.56  &    168.81  &    14.8   &  \phn614  \\
 KOI-4054.01  &     \nodata            &      6428794  &  14.566  &    +15.14  &    169.13  &    16.7   &  \phn641  \\
 KOI-4356.01  &   Kepler-1593\,b (3)   &      8459663  &  15.873  & \phn+7.30  &    174.51  &    11.0   &     1530  \\
 KOI-4450.01  &   Kepler-1606\,b (3)   &      7429240  &  15.139  &    +15.75  &    196.44  &    11.1   &  \phn632  \\
 KOI-4550.01  &   Kepler-1653\,b (4)   &      5977470  &  15.429  & \phn+8.25  &    140.25  & \phn9.6   &  \phn568  \\
 KOI-5236.01  &     \nodata            &      6067545  &  13.093  & \phn+7.01  &    550.86  &    12.1   &  \phn346  \\
 KOI-5856.01  &   Kepler-1638\,b (3)   &     11037818  &  14.759  &    +12.29  &    259.34  &    10.9   &  \phn350 
\enddata

\tablecomments{Columns after the first indicate the \kepler\ planet
  designation and source, \kepler\ identification number, brightness
  in the \kepler\ passband, Galactic latitude, orbital period,
  multiple-event statistic, and transit depth in parts per million
  relative to the out-of-transit stellar flux. For consistency in this
  paper we will refer to all objects by their original KOI names
  throughout. The sources of the \kepler\ planet designations are
  the following: (1) \cite{Barclay:2013}; (2) \cite{Rowe:2014}; (3)
  \cite{Morton:2016}; (4) new designation based on this work.}

\end{deluxetable*}

\section{Photometry}
\label{sec:photometry}

\kepler\ photometry for all targets was retrieved from the Mikulski
Archive for Space Telescopes (MAST)\footnote{\url
    https://archive.stsci.edu/index.html}. The observations used
here are those labeled in the FITS files as PDC\_FLUX, based on Data
Release\;25.  Our targets have a mix of Long Cadence (LC) observations
with $\sim$30\;min samples and Short Cadence (SC) observations with
$\sim$1\;min sampling, and SC observations took precedence when both
LC and SC were available for any observation window.

The PDC photometry adopted for our analysis includes corrections for
instrumental trends as well as estimates of dilution due to other
stars that contaminate the photometric aperture \citep{Stumpe:2014}.
Additionally, we accounted for further dilution from close companions
to our targets identified via high-resolution imaging (see
Section\;\ref{sec:followup}), both in our false positive assessments
(Section\;\ref{sec:validation}) and in our transit light-curve
modeling (Section\;\ref{sec:fits}).  We did not rescale the PDC
photometry based on additional information regarding aperture
dilution.
                      
To minimize the impact of stellar variability on our transit models
described later we applied a polynomial filter to remove variability
on timescales longer than 5 days, as described in Sect.\;4 of
\citet{Rowe:2014}.  A bandpass of 5 days was chosen to be
significantly longer than longest observed transit duration in our
sample, which is 15.9 hours for KOI-5856.01 (Kepler-1638\,b).

\section{Follow-up observations and centroid motion analysis}
\label{sec:followup}

\subsection{High-resolution imaging}
\label{sec:imaging}

Each of our targets has been subjected to high spatial resolution
imaging to detect close stellar companions that could be the source of
the transit signals we observe, if those companions are eclipsed by
another body.  Even if they are not eclipsing and the planet orbits
the KOI, these companion stars can still attenuate the transit signal
and make the planet appear smaller than it really is.

\cite{Furlan:2017} have recently published a compilation of all the
information from the imaging observations performed on more than 1900
stars observed by \kepler. For our sample we have used images gathered
mainly with the following instruments and telescopes: adaptive optics
observations on the Keck\,II 10m telescope with the NIRC2 instrument
\citep{Wizinowich:2004} in the $J$ band (1.246\,$\mu$m) or
$K^{\prime}$ band centered on Br$\gamma$ (2.18\,$\mu$m); adaptive
optics on the 5m Palomar telescope with the PHARO instrument
\citep{Hayward:2001} in $H$ (1.635\,$\mu$m) or $K_s$ (2.145\,$\mu$m);
adaptive optics with the Robo-AO system on the Palomar 1.5m telescope
\citep{Baranec:2016} using a long-pass filter (LP600) starting at
600\,nm, making it similar to the \kepler\ passband; speckle
interferometry on the Gemini-N 8m telescope and on the DCT 4m
telescope with the DSSI instrument \citep{Horch:2009, Horch:2010}
obtained at wavelengths of 562\,nm (Gemini-N only), 692\,nm, or
880\,nm; and in two cases we used imaging with the F555W filter of
WFC3 on the {\it HST} as reported by \cite{Gilliland:2015}.

A total of eight of our KOIs have close companions reported by
\cite{Furlan:2017} within 4\arcsec, occasionally more than one per
target. They are listed in Table\;\ref{tab:closecompanions} along with
the relative position and brightness difference compared to the
primary star. Some of those companions are close enough that they
cannot be ruled out as potential sources of the signal by the centroid
motion analysis described in the next section, and are flagged in the
table. We discuss them further in Section~\ref{sec:validation}.

\begin{deluxetable*}{lccl}
\tablewidth{0pc}
\tablecaption{Close Companions from High-Resolution Imaging.\label{tab:closecompanions}}
\tablehead{
\colhead{Star} &
\colhead{$\rho$ (\arcsec)\tablenotemark{a}} &
\colhead{P.A.\ (deg)} &
\colhead{Brightness difference (mag) and passband}
}
\startdata
KOI-0438  &  3.290~$\pm$~0.059  &  182.0~$\pm$~1.2  &  3.11~$\pm$~0.04 (LP600), 2.245~$\pm$~0.010 ($J$), 2.160~$\pm$~0.010 ($K$) \\
KOI-0854  &  \phm{*}0.016~$\pm$~0.050*  &  209.4~$\pm$~1.0  &  0.299~$\pm$~0.231 ($K$) \\
KOI-0854  &  \phm{*}0.154~$\pm$~0.050*  &  181.6~$\pm$~1.0  &  3.589~$\pm$~0.076 ($K$) \\
KOI-2418  &  \phm{*}0.108~$\pm$~0.050*  &  \phn\phn3.2~$\pm$~1.6    &  3.22~$\pm$~0.15 (692\,nm), 2.94~$\pm$~0.15 (880\,nm), 2.509~$\pm$~0.062 ($K$) \\
KOI-2418  &  2.387~$\pm$~0.050  &  104.4~$\pm$~1.0  &  7.793~$\pm$~0.086 ($K$) \\
KOI-2418  &  3.918~$\pm$~0.066  &  329.2~$\pm$~1.5  &  5.45~$\pm$~0.16 ($J$), 6.845~$\pm$~0.035 ($K$) \\
KOI-2626  &  \phm{*}0.164~$\pm$~0.050*  &  183.4~$\pm$~3.4  &  1.646~$\pm$~0.094 (F555W), 1.302~$\pm$~0.094 (F775W), 1.95~$\pm$~0.15 (562\,nm), \\
          &                     &                   &  2.22~$\pm$~0.15 (692\,nm), 1.28~$\pm$~0.15 (880\,nm) 0.91~$\pm$~0.19 ($K$) \\
KOI-2626  &  \phm{*}0.206~$\pm$~0.050*  &  212.7~$\pm$~1.4  &  0.763~$\pm$~0.071 (F555W), 0.509~$\pm$~0.071 (F775W), 1.91~$\pm$~0.15 (562\,nm), \\
          &                     &                   &  1.63~$\pm$~0.15 (692\,nm), 0.88~$\pm$~0.15 (880\,nm) 0.464~$\pm$~0.079 ($K$) \\
KOI-2650  &  3.121~$\pm$~0.050  &  124.8~$\pm$~1.0  &  8.200~$\pm$~0.059 (F555W), 7.577~$\pm$~0.031 (F775W), 7.252~$\pm$~0.079 ($K$) \\
KOI-3010  &  \phm{*}0.334~$\pm$~0.050*  &  304.5~$\pm$~1.1  &  0.595~$\pm$~0.050 (F555W), 0.294~$\pm$~0.050 (F775W), 0.74~$\pm$~0.15 (692\,nm), \\
          &                     &                   &  0.01~$\pm$~0.15 (880\,nm), 0.245~$\pm$~0.052 ($K$) \\
KOI-3497  &  \phm{*}0.795~$\pm$~0.064*  &  174.1~$\pm$~1.8  &  1.23~$\pm$~0.12 (LP600), 1.31~$\pm$~0.47 ($K$) \\
KOI-4550  &  \phm{*}1.045~$\pm$~0.069*  &  143.7~$\pm$~1.3  &  0.040~$\pm$~0.020 (LP600), 0.75~$\pm$~0.15 (692\,nm), 1.11~$\pm$~0.15 (880\,nm), \\
          &                     &                   &  0.578~$\pm$~0.010 ($J$) \\
KOI-5236  &  \phm{*}1.943~$\pm$~0.050*  &  283.4~$\pm$~1.0  &  6.398~$\pm$~0.028 ($K$) 
\enddata

\tablenotetext{a}{Companions flagged with an asterisk are inside the
  limits of the 3$\sigma$ exclusion region from the centroid motion
  analysis (Table\;\ref{tab:centroids}), and thus cannot be rejected as
  potential false positive sources in that way.}

\end{deluxetable*}

\begin{figure}
\epsscale{1.10}
\plotone{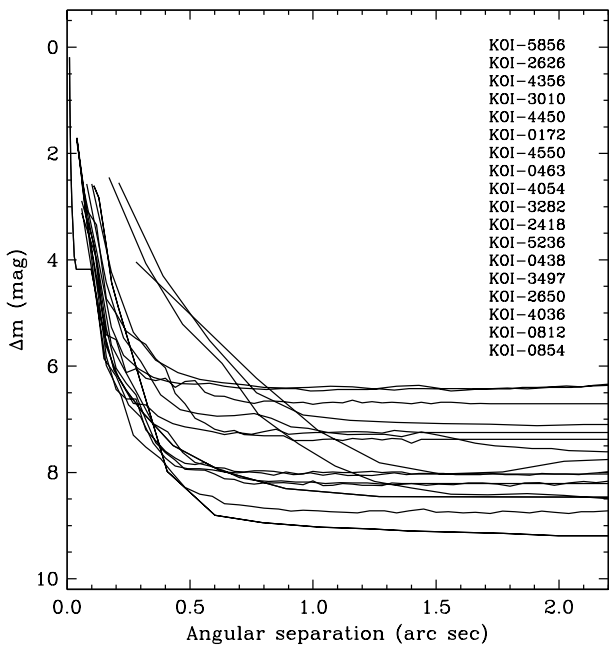}
                                                                                                                                   
\figcaption[]{Sensitivity curves for each of our targets from the
  high-resolution imaging observations that provide the strongest
  constraints on unseen companions. Curves correspond to the KOIs as
  listed, top to bottom, at a reference separation of 2\arcsec. The
  measurements shown are based on Keck AO images in the
  (2.124\,$\mu$m) $K^\prime$ band, with the exception of KOI-0812 and
  KOI-0854, for which we show curves for the F555W passband of {\it
    HST} as measured by
  \cite{Gilliland:2015}. \label{fig:sensitivity}}

\end{figure}

For each star in our sample we have also estimated our ability to
detect companions at close separations. This information is critical
to our analysis described later to validate the candidates.  When not
already available on the Exoplanet Follow-up Observing Program
(ExoFOP) web page\footnote{\url
  https://exofop.ipac.caltech.edu/cfop.php}, we used numerical
simulations to produce sensitivity curves of detectable companion
brightness as a function of angular separation as described by
\cite{Torres:2015} and \cite{Furlan:2017}, or adopted published
sensitivity curves in the case of the Robo-AO and {\it HST}
observations. In two cases we adopted published sensitivity curves for
the Keck observations by \cite{Kraus:2016}.
Figure\;\ref{fig:sensitivity} shows the curves that provide the
strongest constraints on unseen companions for each KOI.  We note that
for several of our targets we used sensitivity curves from several
instruments or at several wavelengths at the same time, to maximize
our ability to rule out blends.

\subsection{Centroid motion}
\label{sec:centroids}

The positions of the flux centroids of our stars offer an additional
and powerful discriminant against false positives, and can provide
evidence that the source of the signal is not on the target if the
image centroid moves significantly in and out of transit. The Data
Validation Reports from the \kepler\ Science Operations Center
pipeline, available also on the NASA Exoplanet Archive, list a measure
of this offset for each quarter computed by subtracting the best-fit
Pixel Response Function centroid out of transit from the difference
image centroid \citep[see][]{Bryson:2013}, as well as a multi-quarter
average and its uncertainty. If the source of a transit is located
less than 3$\sigma$ from the host star, it is considered to be
statistically indistinguishable from the host star. This 3$\sigma$
exclusion radius is mainly a function of the signal-to-noise ratio of
the transit signal.

\begin{deluxetable}{cccc}
\tablewidth{0pc}
\tablecaption{Centroid Results for the Candidates.\label{tab:centroids}}
\tablehead{
\colhead{} &
\colhead{Offset from} &
\colhead{Offset from} &
\colhead{3$\sigma$ exclusion}
\\
\colhead{Candidate} &
\colhead{star (\arcsec)} &
\colhead{star ($\sigma$)} &
\colhead{radius (\arcsec)}
}
\startdata
KOI-0172.02  &  $1.46 \pm  0.59$  &    2.47  &   1.77  \\
KOI-0438.02  &  $0.26 \pm  0.18$  &    1.48  &   0.54  \\
KOI-0463.01  &  $0.03 \pm  0.08$  &    0.35  &   0.24  \\
KOI-0812.03  &  $0.08 \pm  0.20$  &    0.37  &   0.64  \\
KOI-0854.01  &  \phm{\tablenotemark{a}}$0.35 \pm  0.38$\tablenotemark{a}  &   0.93  &   \phm{\tablenotemark{a}}1.13\tablenotemark{a}  \\
KOI-2418.01  &  $0.27 \pm  0.28$  &    0.97  &   0.84  \\
KOI-2626.01  &  \phm{\tablenotemark{a}}$0.62 \pm  0.86$\tablenotemark{a}  &   0.72  &   \phm{\tablenotemark{a}}2.57\tablenotemark{a}  \\
KOI-2650.01  &  $0.53 \pm  0.57$  &    0.94  &   1.70  \\
KOI-3010.01  &  \phm{\tablenotemark{a}}$0.62 \pm  1.58$\tablenotemark{a}  &   0.39  &   \phm{\tablenotemark{a}}4.73\tablenotemark{a}  \\
KOI-3282.01  &  $0.14 \pm  0.33$  &    0.42  &   0.99  \\
KOI-3497.01  &  \phm{\tablenotemark{a}}$0.14 \pm  0.45$\tablenotemark{a}  &   0.32  &   \phm{\tablenotemark{a}}1.34\tablenotemark{a}  \\
KOI-4036.01  &  $0.62 \pm  0.34$  &    1.80  &   1.03  \\
KOI-4054.01  &  $0.32 \pm  0.32$  &    1.01  &   0.95  \\
KOI-4356.01  &  $0.99 \pm  0.58$  &    1.71  &   1.73  \\
KOI-4450.01  &  $0.57 \pm  0.42$  &    1.35  &   1.28  \\
KOI-4550.01  &  \phm{\tablenotemark{a}}$1.17 \pm  1.26$\tablenotemark{a}  &   0.93  &   \phm{\tablenotemark{a}}3.79\tablenotemark{a}  \\
KOI-5236.01  &  $1.07 \pm  0.99$  &    1.09  &   2.96  \\
KOI-5856.01  &  $0.59 \pm  0.32$  &    1.84  &   0.97  

\enddata

\tablenotetext{a}{Offset uncertainties and corresponding 3$\sigma$
  exclusion regions conservatively doubled in size to account for
  possible biases due to the presence of close and relatively bright
  companions.}

\end{deluxetable}

In Table\;\ref{tab:centroids} we list the in-and-out-of-transit offsets
and their uncertainties as reported for each of our targets in the
NASA Exoplanet Archive, in seconds of arc and also in units of the
uncertainties ($\sigma$). The last column gives the corresponding
3$\sigma$ exclusion radii. In all cases the KOIs have offsets that are
less than three times their uncertainties, thus suggesting the targets
are the source of the signals. These exclusion regions are small
enough that they rule out some of the wider companions reported in the
previous section, but not all. We note, on the other hand, that the
centroid offsets in Table\;\ref{tab:centroids} were computed assuming
that the host stars do not have close companions of similar brightness
($\rho < 4\arcsec$, $\Delta m \lesssim 2$~mag). However, as seen in
Table\;\ref{tab:closecompanions} several of our targets do have such
companions, which can introduce error in the centroid measurements. To
be conservative we have doubled the size of the offset uncertainties,
and therefore of the 3$\sigma$ exclusion regions for those
cases. These instances are indicated in Table\;\ref{tab:centroids}.

\subsection{Spectroscopic observations}
\label{sec:spectroscopy}

Many of our targets have been observed spectroscopically as part of
the Exoplanet Follow-up Observing Program for \kepler. A variety of
instruments and telescopes have been used, including HIRES
\citep{Vogt:1994} on the 10m Keck~I telescope on Mauna Kea (Hawaii)
with a typical resolving power of $R \approx 60,000$, TRES
\citep{Furesz:2008} on the 1.5m Tillinghast reflector at the Fred
L.\ Whipple Observatory (Arizona) with $R \approx 44,000$, the Tull
spectrograph \citep{Tull:1995} on the 2.7m McDonald Observatory
telescope (Texas) with $R \approx 60,000$, and the RCSpec instrument
\citep[see][]{Everett:2013} on the 4m Kitt Peak National Observatory
telescope (Arizona) with $R \approx 3000$. The reduced observations
are available on the ExoFOP web page, and we list them in
Table\;\ref{tab:spectra} with their barycentric Julian dates of
observation (BJD), signal-to-noise ratios (SNRs) per resolution
element, and measured heliocentric radial velocities. Additionally,
several of the cooler targets have been observed in the near-infrared
by others, including \cite{Muirhead:2012, Muirhead:2014},
\cite{Mann:2013a, Mann:2013b}, and \cite{Newton:2015}. Details of the
instrumentation and reduction procedures for those observations may be
found there.

\begin{deluxetable*}{lccc@{}ccc@{}ccc@{}ccc@{}c}
\tablewidth{0pc}
\tablecaption{Optical Spectra for our Targets.\label{tab:spectra}}
\tablehead{
\colhead{} &
\multicolumn{2}{c}{Keck/HIRES} &&
\multicolumn{2}{c}{TRES} &&
\multicolumn{2}{c}{McDonald} &&
\multicolumn{2}{c}{KPNO} &&
\colhead{} \\
\cline{2-3} \cline{5-6} \cline{8-9} \cline{11-12} && \\[-1ex]
\colhead{Star} &
\colhead{BJD} &
\colhead{SNR} &&
\colhead{BJD} &
\colhead{SNR} &&
\colhead{BJD} &
\colhead{SNR} &&
\colhead{BJD} &
\colhead{SNR} &&
\colhead{RV (km s$^{-1}$)}
}
\startdata
 KOI-0172 & 2455735.8686 & 58.7    && 2456462.8167  & 26.9    &&   \nodata    & \nodata &&   \nodata    & \nodata &&  $-$41.2 \\
 KOI-0438 & 2456147.9907 & 27.3    &&   \nodata     & \nodata &&   \nodata    & \nodata &&   \nodata    & \nodata &&  $-$11.6 \\
 KOI-0463 & 2456028.1245 & 18.5    &&   \nodata     & \nodata &&   \nodata    & \nodata &&   \nodata    & \nodata &&  $-$96.8 \\
 KOI-0812 & 2456166.0273 & 13.4    &&   \nodata     & \nodata &&   \nodata    & \nodata &&   \nodata    & \nodata &&  $-$30.6 \\
 KOI-0854 & 2456172.8635 & 16.8    &&   \nodata     & \nodata &&   \nodata    & \nodata &&   \nodata    & \nodata &&  \phn$-$3.2 \\
 KOI-2418 & 2456177.9199 & 19.4    &&   \nodata     & \nodata &&   \nodata    & \nodata &&   \nodata    & \nodata &&  \phn$+$5.4 \\
 KOI-2626 & 2456179.8154 & 17.3    &&   \nodata     & \nodata &&   \nodata    & \nodata &&   \nodata    & \nodata &&  \phn$-$2.5 \\
 KOI-2650 & 2456166.9364 & 17.7    &&   \nodata     & \nodata &&   \nodata    & \nodata &&   \nodata    & \nodata &&  $-$22.4 \\
 KOI-3010 & 2456208.8892 & 19.5    &&   \nodata     & \nodata &&   \nodata    & \nodata &&   \nodata    & \nodata &&  \phn$+$0.7 \\
 KOI-3282 & 2456880.9972 & 13.6    &&   \nodata     & \nodata &&   \nodata    & \nodata &&   \nodata    & \nodata &&  \phn$-$2.1 \\
 KOI-3497 & 2456508.0625 & 90.0    &&   \nodata     & \nodata &&   \nodata    & \nodata &&   \nodata    & \nodata &&  $-$17.9 \\
 KOI-4036 & 2456588.8874 & 46.7    &&   \nodata     & \nodata &&   \nodata    & \nodata &&   \nodata    & \nodata &&  $-$20.7 \\
 KOI-4054 & 2457265.8981 & 36.9    &&   \nodata     & \nodata &&   \nodata    & \nodata &&   \nodata    & \nodata &&  $-$25.6 \\
 KOI-4356 & 2456882.0965 & 12.9    &&   \nodata     & \nodata &&   \nodata    & \nodata &&   \nodata    & \nodata &&  \phn$+$8.6 \\
 KOI-4450 &  \nodata     & \nodata &&   \nodata     & \nodata && 2456589.6857 & 16.9    && 2456536.7629 & 32.0    &&  $-$44.9 \\
 KOI-4550 &  \nodata     & \nodata &&   \nodata     & \nodata &&   \nodata    & \nodata && 2456536.7130 & 26.5    && \nodata \\
 KOI-5236 & 2457265.0439 & 56.3    &&  2456934.6175 & 32.6    &&   \nodata    & \nodata &&   \nodata    & \nodata &&  $-$31.5 \\
 KOI-5856 &  \nodata     & \nodata &&   \nodata     & \nodata && 2456814.7854 & 23.7    &&   \nodata    & \nodata &&  $-$94.3 
\enddata
\end{deluxetable*}

All of the optical and near-infrared spectra appear single-lined. To
aid in ruling out potential false positives that could be causing the
transit signals, the 15 targets with available Keck/HIRES observations
(i.e., all except for KOI-4450, 4550, and 5856) were subjected to
injection/recovery simulations to determine limits to the brightness
of potential stellar companions that may fall within the slit of the
HIRES instrument, which has a typical half width of 0\farcs43. For
details of the procedure we refer the reader to the work of
\cite{Kolbl:2015}. We find that in general we should be able to detect
companions brighter than about 1\% of the flux of the target star
($\Delta m = 5$~mag) provided the line separation is more than
10~\kms\ in velocity space. For smaller separations we assume the line
blending would prevent detection of any companions.  For the coolest
stars and for targets with lower SNR spectra the limits are not as
strong, and we are only able to detect companions down to about 3\% of
the flux of the primary star ($\Delta m = 3.8$~mag) at velocity
separations larger than 20~\kms. This is the case for KOI-0172, 0463,
2626, 2650, 3282, 3497 (10~\kms\ limit), and 4356. For KOI-4450, 4550,
and 5856 we adopted a brighter threshold corresponding to $\Delta m =
2$~mag (16\% relative flux), and assumed that any spectroscopic
companions would escape detection if their radial velocity is within
30~\kms\ of that of the main target.

\section{Stellar properties}
\label{sec:stellarproperties}

The atmospheric properties of the host stars including the effective
temperature, surface gravity, and metallicity ($T_{\rm eff}$, $\log
g$, [Fe/H]) are key inputs to infer their absolute mass and radius,
which we describe below. Here we have derived the atmospheric
properties using several methods depending on the spectra
available. For the Keck/HIRES spectra we applied two different
procedures. One is a version of the {\tt SpecMatch-Emp} algorithm
\citep{Yee:2017}\footnote{\url
  https://github.com/samuelyeewl/specmatch-emp} that compares an
observed spectrum against a large library of observed spectra of
well-characterized stars obtained with the same instrument, and finds
the best match. This procedure works well for solar-type stars, but
degrades below $T_{\rm eff} \sim 4300$~K because such stars are not as
well represented in the reference library used for this work. The
other procedure applied to the Keck observations is {\tt SPC}
\citep{Buchhave:2012}, which uses a cross-correlation technique to
compare a region centered on the \ion{Mg}{1}~b triplet
($\sim$5187~\AA) in the observed spectrum against a grid of synthetic
templates based on model atmospheres by R.\ L.\ Kurucz. The same {\tt
  SPC} procedure was applied to the TRES spectra, although the smaller
telescope and use of a narrow wavelength range limited this to the
brighter stars. Results for the McDonald Observatory spectra were
obtained using the {\tt Kea} analysis tool \citep{Endl:2016}, and the
KPNO spectra were analyzed with the procedures described by
\cite{Everett:2013}. Both of these methods rely on libraries of
synthetic spectra.

For stars hotter than 4300~K we have taken an average of all available
results and we report them in Table\;\ref{tab:teffloggmet} along with
the sources. For KOI-5856 the parameters resulting from the {\tt Kea}
analysis provide a less satisfactory fit to the McDonald spectrum than
those reported by \cite{Huber:2014}; we have therefore adopted the
latter, along with their considerably larger formal
  uncertainties. For KOI-0172 the {\tt SpecMatch-Emp} analysis
converged on a subgiant classification ($\log g = 3.82$) as well as a
significantly cooler temperature and lower metallicity than two other
determinations that are based on spectra from two different telescopes
favoring a main-sequence classification. We have adopted the
main-sequence results.

\begin{deluxetable*}{lccccccccc}
\tablewidth{0pc}
\tablecaption{Spectroscopic Properties and Age Constraints for the Targets.\label{tab:teffloggmet}}
\tablehead{
\colhead{} &
\colhead{$T_{\rm eff}$} &
\colhead{[Fe/H]} &
\colhead{$\log g$} &
\colhead{$P_{\rm rot}$\tablenotemark{a}} &
\colhead{Gyro age} &
\colhead{$T_{\rm eff}$} &
\colhead{[Fe/H]} &
\colhead{$\log g$} &
\colhead{$P_{\rm rot}$}
\\
\colhead{Star} &
\colhead{(K)} &
\colhead{(dex)} &
\colhead{(cm s$^{-2}$)} &
\colhead{(days)} &
\colhead{(Gyr)} &
\colhead{Source} &
\colhead{Source} &
\colhead{Source} &
\colhead{Source}
}
\startdata
KOI-0172                  & $5700 \pm 100$      &  $-0.10 \pm 0.10$        &  $4.34 \pm 0.10$        &  \nodata   &      \nodata       &  1,3    &  1,3   &  1,3    &  \nodata     \\
KOI-0438                  & $4057^{+251}_{-64}$    &  $-0.33 \pm 0.11$        &    \nodata              &   26.43    &  $2.9^{+0.3}_{-0.3}$  &   6     &   6    & \nodata &  11,13,15    \\
KOI-0463                  & $3430 \pm 48$       &  $-0.25 \pm 0.06$        &    \nodata              &   50.47    &  $6.5^{+1.9}_{-1.2}$  & 7,9,10  &  7,8   & \nodata &   12,13      \\
KOI-0812                  & $3996 \pm 55$       &  $-0.51 \pm 0.06$        &    \nodata              &   39.49    &  $5.8^{+0.5}_{-0.2}$  &  7,9    &  7,8   & \nodata &   11,13      \\
KOI-0854                  & $3674 \pm 40$       &  $+0.21 \pm 0.05$        &    \nodata              &   20.09    &  $1.4^{+0.7}_{-0.3}$  & 7,9,10  &  7,8   & \nodata & 11,12,13,15  \\
KOI-2418                  & 3577 $\pm$ 58       &  +0.16 $\pm$ 0.08        &    \nodata              &   17.63    &  $1.2^{+0.7}_{-0.3}$  &   9     &   8    & \nodata & 11,12,13,15  \\
KOI-2626                  & $3638 \pm 51$       &  $-0.30 \pm 0.12$        &    \nodata              &   31.18    &  $3.2^{+0.7}_{-0.8}$  &  7,9    &  7,8   & \nodata &    11        \\
KOI-2650                  & $3931 \pm 47$       &  $-0.02 \pm 0.09$        &    \nodata              &   19.92    &  $1.5^{+0.6}_{-0.3}$  & 7,9,10  &  7,8   & \nodata & 11,12,13,15  \\
KOI-3010                  & $3950 \pm 56$       &  $+0.04 \pm 0.12$        &    \nodata              &   14.09    &  $0.8^{+0.6}_{-0.2}$  &  7,9    &   7    & \nodata & 11,13,14,15  \\
KOI-3282                  & $3950 \pm 113$      &  $-0.21 \pm 0.11$        &    \nodata              &   18.32    &  $1.3^{+0.5}_{-0.2}$  &  7,10   &   7    & \nodata &   12,14,15   \\
KOI-3497                  & $3484 \pm 91$       &  $+0.15 \pm 0.11$        &    \nodata              &  $\sim$10  &  $1.0^{+1.0}_{-0.5}$  &   7     &   7    & \nodata &    16        \\ [0.5ex]
KOI-4036                  & $4820 \pm 100$      &  $-0.08 \pm 0.10$        &  $4.54 \pm 0.10$        &  \nodata   &      \nodata       &  1,2    &  1,2   &  1,2    &  \nodata     \\ [0.5ex]
KOI-4054                  & $5276 \pm 100$      &  $+0.06 \pm 0.10$        &  $4.62 \pm 0.10$        &  \nodata   &      \nodata       &  1,2    &  1,2   &  1,2    &  \nodata     \\ [0.5ex]
KOI-4356                  & $4745 \pm 100$      &  $-0.10 \pm 0.10$        &  $4.54 \pm 0.10$        &  \nodata   &      \nodata       &   2     &   2    &   2     &  \nodata     \\ [0.5ex]
KOI-4450                  & $5461 \pm 100$      &  $+0.12 \pm 0.10$        &  $4.45 \pm 0.15$        &  \nodata   &      \nodata       &   4     &   4    &   4     &  \nodata     \\ [0.5ex]
KOI-4550                  & $4807 \pm 100$      &  $-0.18 \pm 0.10$        &  $4.45 \pm 0.15$        &  \nodata   &      \nodata       &   4     &   4    &   4     &  \nodata     \\ [0.5ex]
KOI-5236                  & $5995 \pm 50$       &  $-0.23 \pm 0.08$        &  $4.30 \pm 0.10$        &  \nodata   &      \nodata       &  1,2,3  &  1,2,3 &  1,2,3  &  \nodata     \\
KOI-5856                  & $5906^{+183}_{-147}$   & $-0.76^{+0.32}_{-0.26}$     &  $4.47^{+0.10}_{-0.29}$     &  \nodata   &      \nodata       &   5     &   5    &   5     &  \nodata  
\enddata

\tablenotetext{a}{Uncertainties are assumed to be 5\%.}

\tablecomments{Sources in the last four columns are:
(1) Keck/HIRES spectrum analyzed with {\tt SPC};
(2) Keck/HIRES spectrum analyzed with {\tt SpecMatch-Emp};
(3) TRES spectrum analyzed with {\tt SPC};
(4) KPNO spectrum analyzed as described by \cite{Everett:2013};
(5) \cite{Huber:2014};
(6) \cite{Muirhead:2012};
(7) \cite{Muirhead:2014};
(8) \cite{Mann:2013a};
(9) \cite{Mann:2013b};
(10) \cite{Newton:2015};
(11) \cite{Walkowicz:2013};
(12) \cite{McQuillan:2013a};
(13) \cite{McQuillan:2013b};
(14) \cite{Nielsen:2013};
(15) \cite{Reinhold:2013};
(16) See text.
}
\end{deluxetable*}

For the cooler stars we have preferred to rely on results from the
literature based on near-infrared spectra, as the techniques based on
optical spectra tend to break down for late K and M stars. Infrared
determinations of $T_{\rm eff}$ and [Fe/H] use various spectral
indices and calibrations developed in recent years
\citep[see][]{Rojas-Ayala:2012, Terrien:2012, Mann:2013a, Mann:2013b,
  Newton:2015}, although there are still some differences between
different methods. For example, \cite{Mann:2013b} indicate their
temperatures are on average 72~K hotter than those of
\cite{Muirhead:2012}, which use the \cite{Rojas-Ayala:2012}
calibration. On the other hand, \cite{Newton:2015} claim their
temperatures are essentially on the same scale as those of
\cite{Mann:2013b} (with only a 10~K difference), which seems to be
confirmed by the fact that both sets of authors claim their respective
scales are some 40~K hotter than those of \cite{Dressing:2013}.  The
latter determinations are photometric rather than spectroscopic, but
can nevertheless serve as a secondary reference. Some of our stars
have been classified also by \cite{Muirhead:2014}, whose temperatures
are on the same scale as those of \cite{Muirhead:2012} given that they
use the same calibrations. To refer all temperatures to the same
scale, we have chosen to adjust the \cite{Muirhead:2012,
  Muirhead:2014} values for our targets by +72~K to place them on the
same scale as the Mann/Newton temperatures. We note also that some of
the results by \cite{Muirhead:2012} have been superseded by those of
\cite{Muirhead:2014}, which used the same spectra.

Several of the above studies report metallicity determinations for the
KOIs in our sample, sometimes in the form of [m/H] indices
\citep{Muirhead:2012}, other times as [Fe/H] indices
\citep{Mann:2013a}, and occasionally both \citep{Muirhead:2014}. These
measures have sometimes been used interchangeably in the literature.
We note, however, that [m/H] and [Fe/H] are not the quite same and are
typically not independent in the above studies.  Published [m/H]
results by these authors are based on a calibration by
\cite{Rojas-Ayala:2012}, and a close look reveals a perfectly linear
relation between that [m/H] calibration and a similar one by the same
authors for [Fe/H]. This is traceable to a similar relation present
between [m/H] and [Fe/H] in the metallicity estimates of many of the
standard stars used by \cite{Rojas-Ayala:2012}, which come from the
work of \cite{Valenti:2005}. Consequently, for this work we have
uniformly used [Fe/H] when both are available from the same source
\citep[as in the study by][]{Muirhead:2014}, and when only [m/H] is
available we have converted it to [Fe/H] using the
\cite{Rojas-Ayala:2012} relations.

For the cool stars in our sample we have adopted weighted averages of
all available near-infrared determinations of $T_{\rm eff}$ and
[Fe/H]. These are reported also in Table\;\ref{tab:teffloggmet}, with
their sources.

A comparison of the spectroscopic properties in
Table\;\ref{tab:teffloggmet} with the values in the catalog of
\cite{Mathur:2017}, which was used for the final transiting planet
search run by the \kepler\ Mission (DR25), indicates a few significant
differences. For KOI-4356 and KOI-5856 our temperatures are
$\sim$400~K hotter and our metallicities differ by +1.4 and $-1.0$~dex
from DR25. For KOI-4356 we suspect this is because the DR25 estimates
are photometric, whereas ours are spectroscopic.  For KOI-5856 we are
not able to trace the source of the DR25 determinations in the
published literature at the time of this writing, and have chosen to
revert to the values reported by \cite{Huber:2014}, as explained
earlier. For KOI-4550 and KOI-5236 our spectroscopic metallicities are
lower by about 0.34~dex, with the DR25 values being from the same
unpublished source indicated above.

Unlike optical determinations for solar-type stars, near-infrared
spectroscopic analyses of cool stars are typically not able provide a
measure of the surface gravity, which has a relatively subtle effect
on the line profiles. Consequently, when attempting to infer the
stellar properties by comparison with model isochrones as described
below, there is no handle on the radii (or age) of these objects,
which then hampers the determination of the planetary radii.  The
sizes of cool stars change relatively little with age, so in the end
this uncertainty is not very important.  Nevertheless, following
\cite{Torres:2015} we have taken advantage of the fact that all of
these cool stars show periodic brightness variations that we interpret
as due to rotational modulation caused by spots. We use the
measured rotation periods ($P_{\rm rot}$) for these stars along with
gyrochronology relations to estimate a rough age, which in turn serves
to constrain the isochrone fits and enables us to infer a radius. The
rotation period estimates are averages from the work of
\cite{Walkowicz:2013}, \cite{McQuillan:2013a, McQuillan:2013b},
\cite{Nielsen:2013}, and \cite{Reinhold:2013}, which agree fairly well
with each other for objects in common. The average periods are listed
with their sources in Table\;\ref{tab:teffloggmet}, and are assumed to
have uncertainties of 5\%. To infer ages we used the gyrochronology
relations by \cite{Epstein:2014}. The required $B-V$ color indices
were adopted from the $UBV$ survey of the \kepler\ field by
\cite{Everett:2012}\footnote{A correction of +0.10~mag was applied to
  the $V$ magnitudes to bring them onto the standard system.}, or from
the AAVSO Photometric All-Sky Survey \citep[APASS;][]{Henden:2012} for
KOI-0438, and were corrected for the presence of the close companions
reported in Section~\ref{sec:imaging} on the assumption that they are
physically associated. Reddening corrections were made using the
reddening values listed in the \kepler\ Input Catalog
\citep[KIC;][]{Brown:2011}.  To account for possible errors in these
adjustments we adopted a conservative uncertainty of 0.10~mag for the
dereddened $B-V$ colors. The ages derived in this way are presented
also in Table\;\ref{tab:teffloggmet}. While there is no published
rotation period for KOI-3497, its raw light curve shows a rather clear
periodicity near 10 days similar to that of Hyades stars of its
spectral type, which suggests a relatively young age. We assign it an
age of $1.0^{+1.0}_{-0.5}$~Gyr with a generous uncertainty.

With these spectroscopic properties and age constraints for the cooler
objects we estimated the physical properties of all our targets (most
importantly the mass and radius) using stellar evolution models from
the Dartmouth series \citep{Dotter:2008}. We employed a Monte Carlo
procedure in which we perturbed each of the three observational
constraints assuming they follow uncorrelated Gaussian distributions
characterized by the errors reported in Table\;\ref{tab:teffloggmet}.
We generated half a million such sets, and for each one we obtained
the best fit to the model isochrones based on a standard $\chi^2$
criterion involving the three observables and their uncertainties. The
analysis was based on a pre-computed grid of models with a metallicity
resolution of 0.01\;dex, sufficient for our purposes, and an age
resolution of 0.5\;Gyr up to 13.5\;Gyr (0.25\;Gyr for ages younger
than 4\;Gyr), with linear interpolation for intermediate age values.
Masses were explored along each isochrone on a fine grid generated by
cubic spline interpolation with a step equal to 1/1000$^{\rm th}$ of
the separation between consecutive mass values in the original
isochrone tables.  Once the best fit was identified, other properties
including the radii, stellar densities, bolometric luminosities, and
passband-specific absolute magnitudes were calculated or simply read
off of the isochrone, with cubic spline interpolation at the best-fit
mass, as needed. We collected the outcome of these simulations into
distributions for the stellar properties of interest, and adopted the
mode of each distribution as a representative value. In all cases the
distributions were unimodal.  The results are presented in
Table\;\ref{tab:physical}, where the uncertainties correspond to the
68.3\% (1$\sigma$) credible intervals. Distances were calculated from
the absolute $K_s$ magnitudes derived from the isochrone fits along
with the apparent brightness from the Two Micron All-Sky Survey
\citep[2MASS;][]{Cutri:2003} in the same passband, which is the one
least affected by interstellar extinction. We made appropriate
corrections for extinction, as well as for the presence of unresolved
companions, as above.

\begin{deluxetable*}{lccccccccc}
\tablewidth{0pc}
\tablecaption{Physical Properties of the Targets Derived from Isochrone Fits.\label{tab:physical}}
\tablehead{
\colhead{} &
\colhead{Age} &
\colhead{$M_{\star}$} &
\colhead{$R_{\star}$} &
\colhead{$\log g$} &
\colhead{$\rho_{\star}$} &
\colhead{$\log L_{\star}$} &
\colhead{$M_V$} &
\colhead{$M_{K_s}$} &
\colhead{Distance}
\\
\colhead{Star} &
\colhead{(Gyr)} &
\colhead{($M_{\sun}$)} &
\colhead{($R_{\sun}$)} &
\colhead{(cm\,s$^{-2}$)} &
\colhead{($\rho_{\sun}$)} &
\colhead{($L_{\sun}$)} &
\colhead{(mag)} &
\colhead{(mag)} &
\colhead{(pc)}
}
\startdata
   KOI-0172 & $ 9.8^{+1.7}_{-4.1}$ & $0.914^{+0.051}_{-0.039}$ & $0.991^{+0.151}_{-0.072}$ & $4.399^{+0.070}_{-0.106}$ & $ 0.85^{+0.31}_{-0.23}$ & $-0.021^{+0.134}_{-0.087}$ & $ 4.89^{+0.23}_{-0.34}$ & $ 3.33^{+0.18}_{-0.31}$ & $ 595^{ +95}_{ -45}$ \\ [+0.5ex]
   KOI-0438 & $ 2.9^{+0.4}_{-0.3}$ & $0.542^{+0.053}_{-0.125}$ & $0.529^{+0.045}_{-0.135}$ & $4.721^{+0.147}_{-0.027}$ & $ 4.11^{+2.73}_{-0.96}$ & $-1.210^{+0.186}_{-0.388}$ & $ 8.78^{+1.28}_{-0.66}$ & $ 5.34^{+0.88}_{-0.39}$ & $ 195^{ +52}_{ -58}$ \\ [+0.5ex]
   KOI-0463 & $ 6.8^{+1.9}_{-1.2}$ & $0.269^{+0.046}_{-0.039}$ & $0.272^{+0.040}_{-0.031}$ & $4.995^{+0.042}_{-0.047}$ & $12.73^{+3.76}_{-2.36}$ & $-2.022^{+0.129}_{-0.143}$ & $11.44^{+0.44}_{-0.37}$ & $ 7.21^{+0.34}_{-0.31}$ & $  69^{ +12}_{  -9}$ \\ [+0.5ex]
   KOI-0812 & $ 6.0^{+0.6}_{-0.3}$ & $0.507^{+0.019}_{-0.023}$ & $0.495^{+0.020}_{-0.026}$ & $4.753^{+0.027}_{-0.018}$ & $ 4.14^{+0.54}_{-0.29}$ & $-1.253^{+0.059}_{-0.074}$ & $ 8.85^{+0.23}_{-0.20}$ & $ 5.48^{+0.17}_{-0.13}$ & $ 331^{ +23}_{ -23}$ \\ [+0.5ex]
   KOI-0854 & $ 1.6^{+0.8}_{-0.3}$ & $0.534^{+0.019}_{-0.022}$ & $0.502^{+0.018}_{-0.021}$ & $4.763^{+0.020}_{-0.015}$ & $ 4.18^{+0.42}_{-0.26}$ & $-1.385^{+0.050}_{-0.060}$ & $ 9.76^{+0.20}_{-0.18}$ & $ 5.62^{+0.14}_{-0.12}$ & $ 321^{ +20}_{ -20}$ \\ [+0.5ex]
   KOI-2418 & $ 1.5^{+0.8}_{-0.2}$ & $0.480^{+0.030}_{-0.046}$ & $0.450^{+0.029}_{-0.041}$ & $4.811^{+0.042}_{-0.026}$ & $5.21^{+1.14}_{-0.58}$ & $-1.533^{+0.084}_{-0.114}$ & $10.24^{+0.33}_{-0.29}$ & $5.98^{+0.28}_{-0.20}$ & $194^{ +21}_{ -22}$ \\ [+0.5ex]
   KOI-2626 & $ 3.2^{+0.7}_{-0.8}$ & $0.404^{+0.040}_{-0.047}$ & $0.382^{+0.036}_{-0.040}$ & $4.878^{+0.046}_{-0.036}$ & $ 7.05^{+1.90}_{-0.98}$ & $-1.638^{+0.093}_{-0.122}$ & $10.22^{+0.35}_{-0.28}$ & $ 6.32^{+0.30}_{-0.23}$ & $ 252^{ +32}_{ -30}$ \\ [+0.5ex]
   KOI-2650 & $ 1.6^{+0.7}_{-0.3}$ & $0.578^{+0.022}_{-0.023}$ & $0.549^{+0.021}_{-0.022}$ & $4.721^{+0.018}_{-0.017}$ & $ 3.47^{+0.32}_{-0.23}$ & $-1.194^{+0.057}_{-0.055}$ & $ 8.85^{+0.20}_{-0.17}$ & $ 5.27^{+0.12}_{-0.15}$ & $ 339^{ +26}_{ -18}$ \\ [+0.5ex]
   KOI-3010 & $ 1.3^{+0.6}_{-0.1}$ & $0.594^{+0.027}_{-0.029}$ & $0.566^{+0.021}_{-0.031}$ & $4.704^{+0.029}_{-0.010}$ & $ 3.25^{+0.43}_{-0.17}$ & $-1.158^{+0.058}_{-0.073}$ & $ 8.79^{+0.23}_{-0.22}$ & $ 5.10^{+0.24}_{-0.07}$ & $ 436^{ +32}_{ -33}$ \\ [+0.5ex]
   KOI-3282 & $ 1.4^{+0.6}_{-0.2}$ & $0.544^{+0.036}_{-0.047}$ & $0.518^{+0.039}_{-0.048}$ & $4.744^{+0.045}_{-0.034}$ & $ 3.89^{+0.89}_{-0.53}$ & $-1.247^{+0.123}_{-0.133}$ & $ 8.94^{+0.46}_{-0.40}$ & $ 5.43^{+0.28}_{-0.32}$ & $ 309^{ +53}_{ -35}$ \\ [+0.5ex]
   KOI-3497 & $ 1.6^{+1.0}_{-0.3}$ & $0.418^{+0.051}_{-0.090}$ & $0.393^{+0.048}_{-0.072}$ & $4.865^{+0.077}_{-0.045}$ & $ 6.76^{+3.18}_{-1.30}$ & $-1.695^{+0.132}_{-0.228}$ & $10.74^{+0.67}_{-0.42}$ & $ 6.37^{+0.55}_{-0.32}$ & $  81^{ +15}_{ -17}$ \\ [+0.5ex]
   KOI-4036 & $ 3.9^{+7.3}_{-0.8}$ & $0.743^{+0.034}_{-0.030}$ & $0.713^{+0.031}_{-0.025}$ & $4.608^{+0.017}_{-0.032}$ & $ 2.06^{+0.17}_{-0.20}$ & $-0.606^{+0.071}_{-0.067}$ & $ 6.64^{+0.24}_{-0.23}$ & $ 4.27^{+0.10}_{-0.12}$ & $ 331^{ +19}_{ -15}$ \\ [+0.5ex]
   KOI-4054 & $ 3.0^{+6.6}_{-0.5}$ & $0.863^{+0.040}_{-0.037}$ & $0.817^{+0.051}_{-0.030}$ & $4.557^{+0.020}_{-0.056}$ & $ 1.58^{+0.16}_{-0.24}$ & $-0.333^{+0.077}_{-0.067}$ & $ 5.76^{+0.21}_{-0.21}$ & $ 3.85^{+0.11}_{-0.15}$ & $ 591^{ +43}_{ -28}$ \\ [+0.5ex]
   KOI-4356 & $ 3.4^{+7.8}_{-0.3}$ & $0.725^{+0.034}_{-0.029}$ & $0.698^{+0.029}_{-0.025}$ & $4.616^{+0.017}_{-0.030}$ & $ 2.14^{+0.18}_{-0.19}$ & $-0.652^{+0.070}_{-0.068}$ & $ 6.80^{+0.25}_{-0.23}$ & $ 4.33^{+0.11}_{-0.11}$ & $ 657^{ +39}_{ -33}$ \\ [+0.5ex]
   KOI-4450 & $ 9.4^{+1.9}_{-5.8}$ & $0.921^{+0.044}_{-0.040}$ & $0.915^{+0.117}_{-0.051}$ & $4.486^{+0.041}_{-0.107}$ & $ 1.19^{+0.22}_{-0.34}$ & $-0.171^{+0.125}_{-0.076}$ & $ 5.30^{+0.22}_{-0.33}$ & $ 3.57^{+0.13}_{-0.28}$ & $ 914^{+126}_{ -58}$ \\ [+0.5ex]
   KOI-4550 & $ 7.7^{+3.6}_{-4.6}$ & $0.719^{+0.035}_{-0.029}$ & $0.693^{+0.030}_{-0.024}$ & $4.618^{+0.017}_{-0.031}$ & $ 2.16^{+0.18}_{-0.20}$ & $-0.636^{+0.071}_{-0.066}$ & $ 6.73^{+0.24}_{-0.23}$ & $ 4.33^{+0.10}_{-0.12}$ & $ 755^{ +44}_{ -36}$ \\ [+0.5ex]
   KOI-5236 & $ 7.6^{+1.1}_{-2.5}$ & $0.959^{+0.037}_{-0.035}$ & $1.075^{+0.174}_{-0.081}$ & $4.335^{+0.087}_{-0.102}$ & $ 0.66^{+0.31}_{-0.16}$ & $+0.130^{+0.131}_{-0.077}$ & $ 4.50^{+0.19}_{-0.33}$ & $ 3.09^{+0.18}_{-0.32}$ & $ 521^{ +86}_{ -39}$ \\ [+0.5ex]
   KOI-5856 & $5.7^{+5.1}_{-2.6}$ & $0.817^{+0.092}_{-0.048}$ & $0.835^{+0.189}_{-0.075}$ & $4.525^{+0.052}_{-0.165}$ & $1.40^{+0.41}_{-0.58}$ & $-0.114^{+0.210}_{-0.120}$ & $5.14^{+0.32}_{-0.54}$ & $3.69^{+0.21}_{-0.48}$ & $764^{+186}_{ -77}$  
\enddata

\tablecomments{Values correspond to the mode of the corresponding
  parameter distributions, with 68.3\% credible intervals (see text).}

\end{deluxetable*}

\section{Candidate validation}
\label{sec:validation}

The faintness of our targets combined with the very small Doppler
signals expected for Earth-size planets in relatively long-period
orbits typically prevents confirmation of their planetary nature by
the radial-velocity technique. The alternative is statistical
validation, whose goal is to show that the likelihood of a true planet
is much greater than that of a false positive. Here we have applied
the \blender\ technique \citep{Torres:2004, Torres:2011, Torres:2015}
that has been used previously to validate many of the most iconic
results from the \kepler\ Mission \citep[see, e.g.,][]{Fressin:2012,
  Borucki:2013, Barclay:2013, Meibom:2013, Kipping:2014, Jenkins:2015,
  Kipping:2016}. \blender\ relies on the information contained in the
shape of the transit light curve to help rule out astrophysical false
positives. The types of false positives considered by
\blender\ include blends with eclipsing binaries as well as with
single stars transited by a (larger) planet. In both cases the
intruding binary system may be in the background, in the foreground,
or even physically bound to the target star. Other ingredients in the
calculations include known distributions of binary star properties
(periods, eccentricities, mass ratios, etc.), the number density of
stars in the vicinity of each KOI from Galactic structure models, and
the estimated rates of occurrence of eclipsing binary stars and
transiting planets as determined from the \kepler\ Mission itself.

As noted in the Introduction, many of the KOIs in our sample have been
validated with other methodologies, and have received official
\kepler\ designations as reported in Table\;\ref{tab:targets}. Three
that are in multi-candidate systems (``multis''), KOI-0438.02,
KOI-0812.03, and KOI-2650.01, were announced as validated by
\cite{Rowe:2014} based on the statistical argument that planet-like
signals in multis have a very high chance of being caused by true
planets \citep{Lissauer:2014}. Ten others received
\kepler\ designations after being validated with the algorithm known
as \vespa\ \citep{Morton:2016}\footnote{\url
  https://github.com/timothydmorton/VESPA}, which adds up the
likelihoods of different types of false positive scenarios and
compares them with that of a planet. \vespa\ has also been applied to
the rest of our targets, though not all have been considered
validated.  In both of these techniques the main false positives
considered are those that involve eclipsing binaries. Scenarios in
which the planet transits an unseen star in the photometric aperture
rather than the intended target, and is therefore larger than it
appears, are not considered to be false positives in those methods as
they are only concerned with demonstrating the presence of a planet
along the line of sight to the target, regardless of its size or exact
location.  The typical threshold for validation in these methods is a
confidence level of 99\%.

Here we are interested in confirming planets that have a chance of
being habitable, so the situation is different in that the size of the
planetary body is just as important as the flux it receives from the
parent star. This is because planets much larger than about
1.8~$R_{\earth}$ are more likely to be gaseous than rocky, and may
lack a solid surface to support liquid water
\citep[e.g.,][]{Rogers:2015, Wolfgang:2016, Fulton:2017}.  Therefore,
the chance that the planet is around a different star than the
intended target, and is thus larger than it appears, must be taken
into account for the validation. As we demonstrate below, these kinds
of false positives tend to dominate the overall blend frequency, and
including them in the analysis makes validations significantly more
challenging. For this reason we consider the previous validations of
most of our targets to be insufficient for our purposes, and we
revisit them with \blender. Additionally, we have chosen to be more
conservative given the importance of these objects and we adopt a
higher threshold for validation corresponding to a 99.73\% confidence
level (3$\sigma$), consistent with previous applications of \blender.

Finally, another of our targets (KOI-0172.02) was considered validated
by \cite{Barclay:2013} at the 99.3\% confidence level using a scheme
somewhat similar to \blender, though with more simplified
assumptions. We revisit this case as well.

Full details of the \blender\ procedure have been described in many of
the previous applications of this technique, as listed at the
beginning of this section, most recently by \cite{Torres:2015}. We
summarize the principles here for the benefit of the reader. For each
KOI \blender\ first simulates a large number of realistic false
positive scenarios involving eclipsing binaries as well as larger
planets transiting blended stars, and compares the synthetic light
curves for each of them against the \kepler\ light curves in a
$\chi^2$ sense.  The properties of the stars involved in blends are
derived based on model isochrones. The simulated transit light curves
account for any extra light in the aperture, such as that coming from
the close companions reported in Section~\ref{sec:imaging}.  Blends
that result in the wrong shape for the transit are considered to be
ruled out. This allows us to place useful constraints on the
properties of the objects making up the blend, including their sizes
and masses, their color and brightness, the line-of-sight distance
between the eclipsing pair and the KOI, the eccentricities of the
orbits, etc.  Monte Carlo simulations are then performed to estimate
the frequencies of blends of different kinds allowed by the above
exploration of parameter space and accounting also for all additional
constraints coming from the follow-up observations (see
below). Finally, the total blend frequency is compared with the
expected frequency of planets of the period and size implied by the
observations (referred to here as the ``planet prior'') to infer the
confidence level of the validation.

To help in the rejection of blends we made use of contrast curves
from the high-resolution imaging for each KOI, which provide
information on the detectability of close companions as a function of
the angular separation from the target. Similarly, our high-resolution
spectroscopy places limits on the brightness of companions falling
within the slit of the spectrograph that would be seen as a second set
of spectral lines, if separated enough from the lines of the target
(Section~\ref{sec:spectroscopy}). The centroid motion analysis is able
to exclude blended objects beyond a certain angular separation (the
3$\sigma$ exclusion regions reported in Section~\ref{sec:centroids}),
regardless of brightness. Finally, we used the observed $r-K_s$ color
of the KOI as listed in the KIC to eliminate blends that are too red
or too blue compared to the target, by more than three times the
observational uncertainties.

\subsection{Results}
\label{sec:blenderresults}

In the following we refer to blend scenarios involving background or
foreground eclipsing binaries as ``BEB''. Blends with a background or
foreground star transited by a larger planet are denoted ``BP''.
Physically associated stars transited by another star or by a planet
(hierarchical triple configurations) are referred to as ``HTS'' and
``HTP'', respectively.

\begin{figure*}
\centering
\begin{tabular}{ccc}
\includegraphics[width=6.3cm]{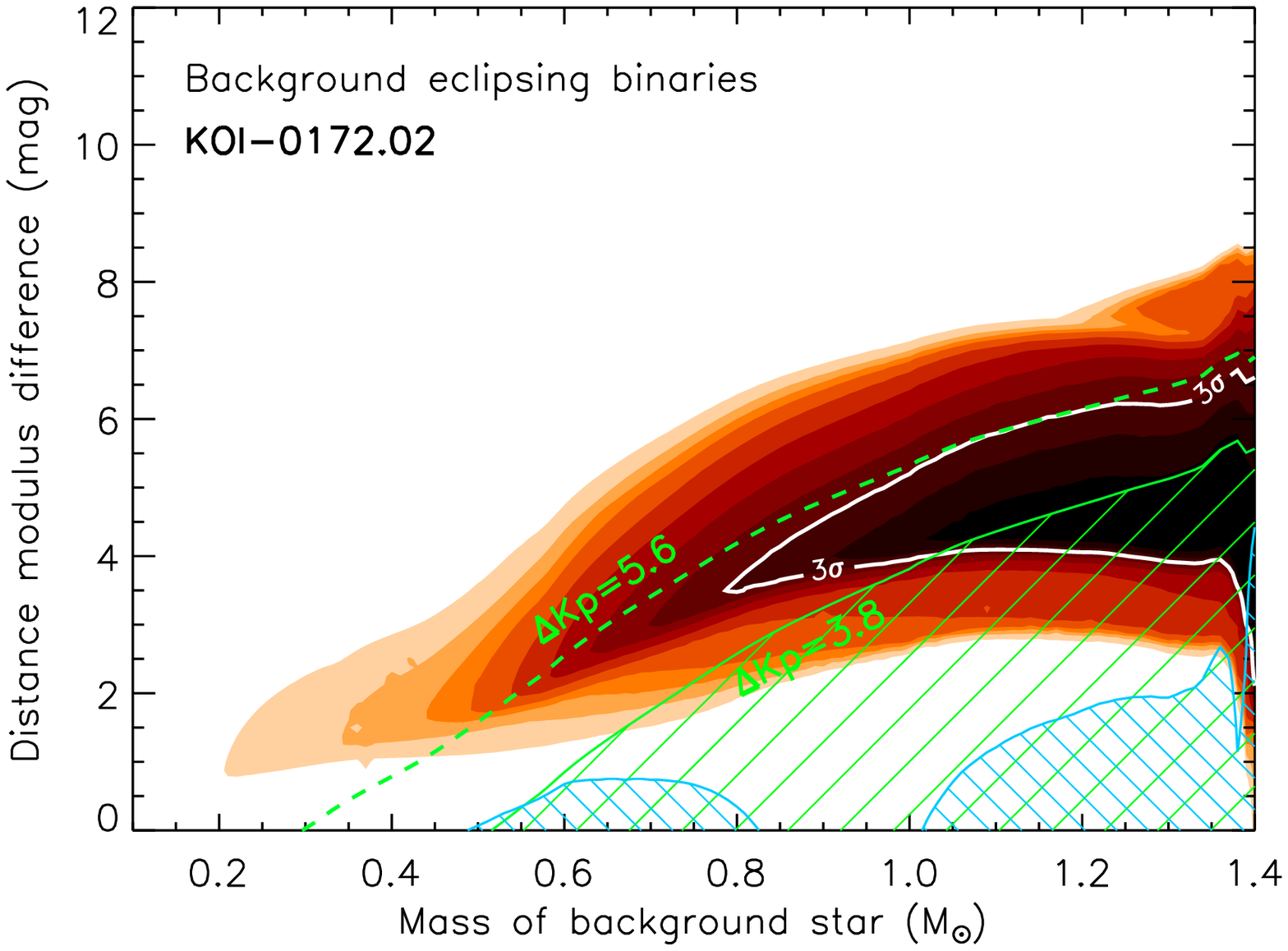} &
\includegraphics[width=6.3cm]{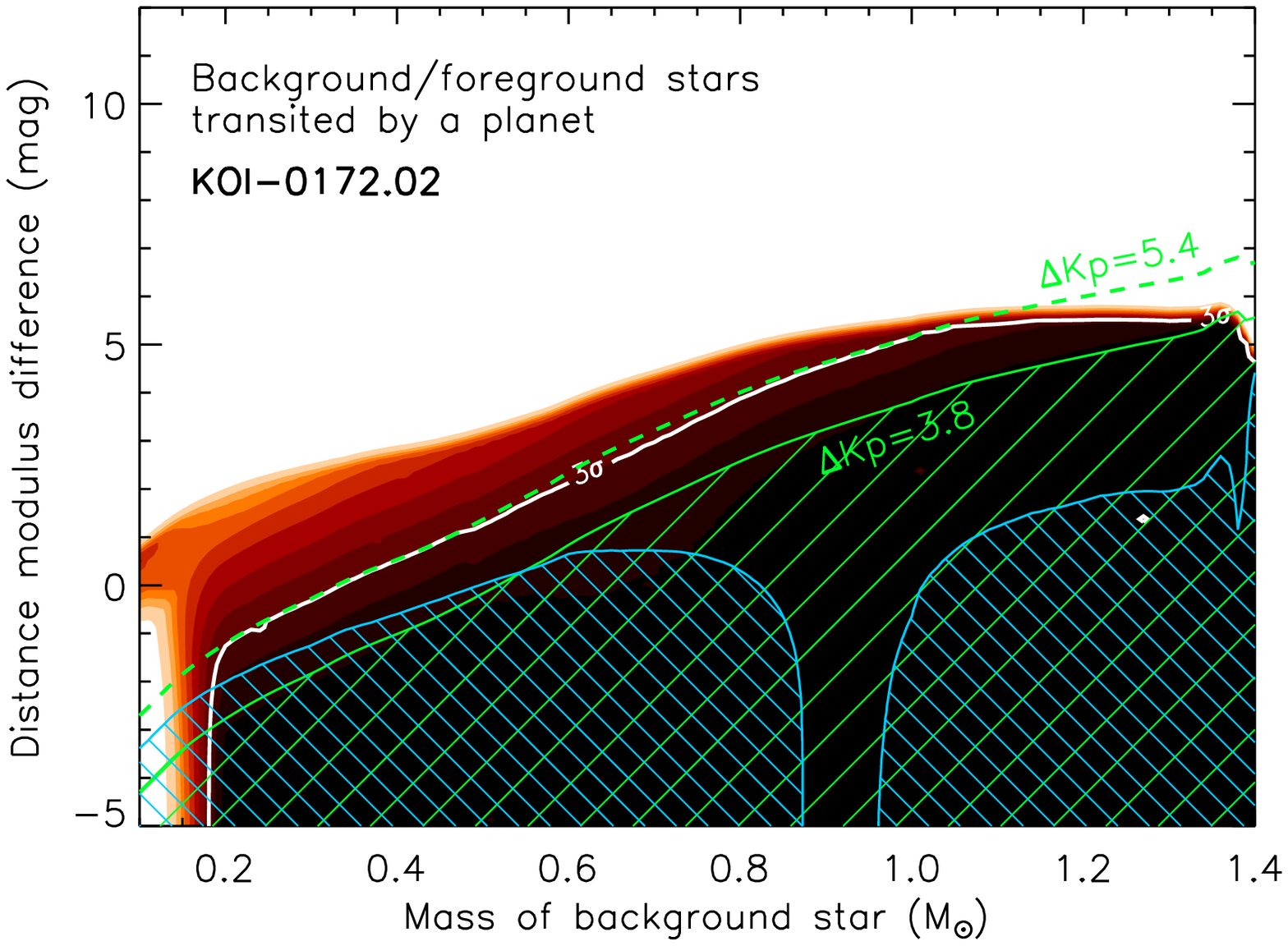} &
\includegraphics[width=6.3cm]{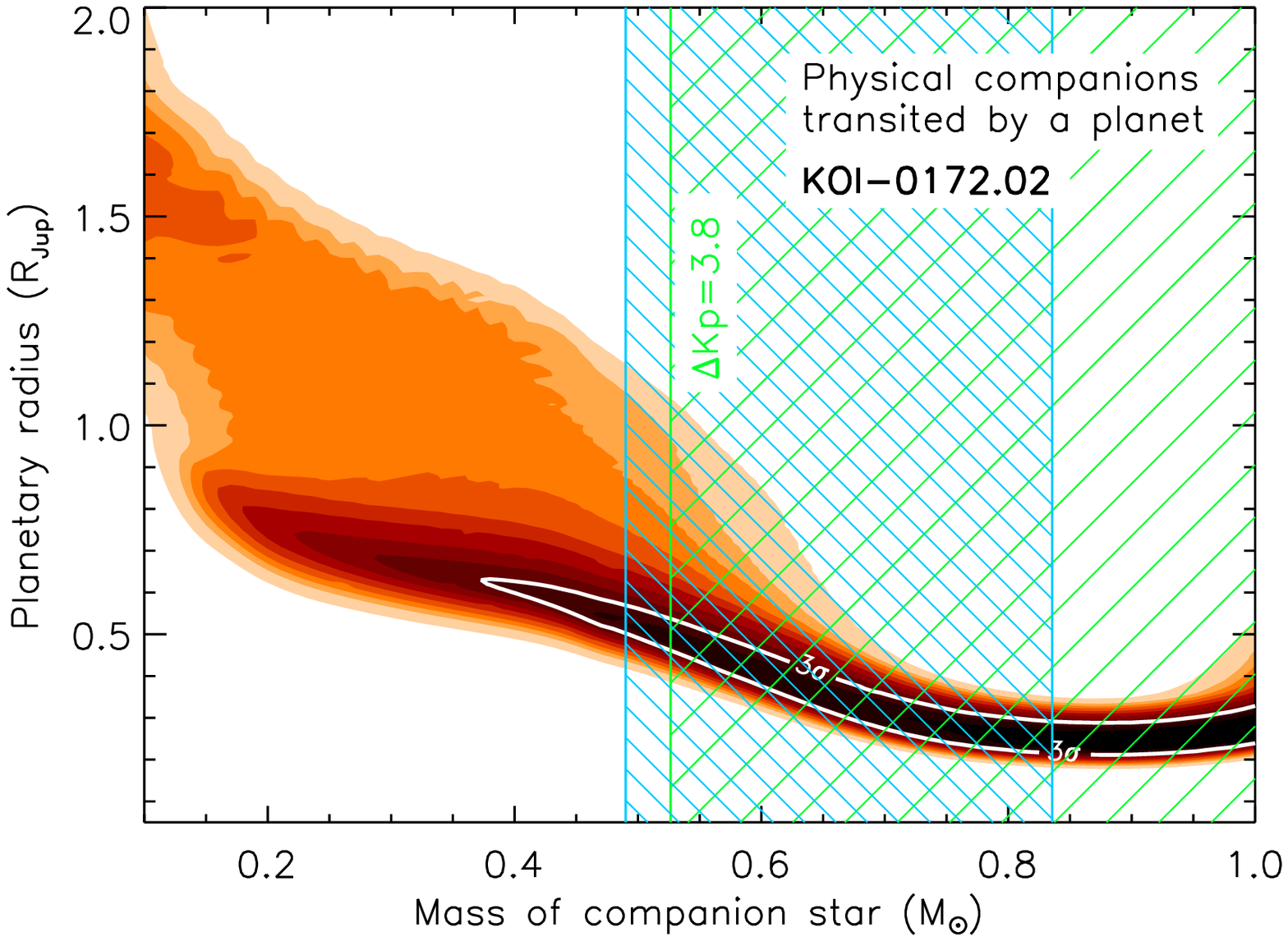} \\
\end{tabular}

\figcaption[]{Map of the $\chi^2$ surface (goodness of fit) for
  KOI-0172.02 for three different blend scenarios, as labeled. Only
  blends within the solid white contours (darker shading) provide fits
  to the \kepler\ light curves that are within acceptable limits
  \citep[3$\sigma$, where $\sigma$ is the significance level of the
    $\chi^2$ difference compared to a transiting planet model fit;
    see][]{Fressin:2012}. Other concentric colored areas (lighter
  colors) represent fits that are increasingly worse (4$\sigma$,
  5$\sigma$, etc.), which we consider to be ruled out.  The hatched
  green areas indicate regions of parameter space where blended stars
  can be excluded if they are within 0\farcs43 of the target
  (half-width of the spectrometer slit), within 3.8 magnitudes in
  brightness (3\% relative flux), and have a radial velocity differing
  from the target by 20~\kms\ or more. In all of those cases they
  would have been detected spectroscopically.  Blends in the hatched
  blue areas can also be ruled out because they would be either too
  red (left) or too blue (right) compared to the measured $r-K_s$
  color of KOI-0172.02, by more than three times the measurement
  uncertainty. \emph{Left:} BEB scenario. The vertical axis represents
  the linear distance between the eclipsing binary and the target
  ($D_{\rm BEB} - D_{\rm targ}$), cast for convenience in terms of the
  distance modulus difference $\Delta\delta = 5 \log(D_{\rm
    BEB}/D_{\rm targ})$. Eclipsing binaries with primary stars above
  the dashed green line ($\Delta K\!p = 5.6$ mag) are too faint to
  mimic the transit. \emph{Middle:} BP scenarios. The combination of
  the $r-K_s$ color constraint and the spectroscopic constraint rules
  out many of the otherwise viable blends. \emph{Right:} HTP
  scenario. The vertical axis now shows the size of the planet
  transiting the companion star, in units of Jupiter's radius. All
  blends of this kind within 3.8~mag of the target are excluded, and
  others are also ruled out for being too blue. Only physical
  companions below 0.49~$M_{\sun}$ can still mimic the \kepler\ light
  curve and go undetected. \label{fig:blender}}

\end{figure*}

We illustrate the constraints derived from the shape of the light
curve and the follow-up observations using KOI-0172.02 as an
example. Figure\;\ref{fig:blender} shows cross-sections of parameter
space for blends from three different scenarios: BEB, BP, and HTP. We
find that no HTS scenarios result in acceptable fits to the
\kepler\ photometry for this candidate, or indeed for any of the other
stars in our sample. For the BEB scenario (left panel),
\blender\ indicates that eclipsing binaries constitute viable blends
(darker region inside the white contour) only if the primary is a
main-sequence star between about 0.8 and 1.4~$M_{\sun}$ (more massive
stars will typically have evolved into a giant, producing the wrong
shape for the transit), and is no fainter in the \kepler\ passband
than 5.6 magnitudes relative to the target. Also shown as hatched
areas are sections of parameter space where blends can be ruled out
based on constraints from spectroscopy (green) and $r-K_s$ color
(blue). The vertical axis represents the distance between the binary
and the target, expressed as a difference in distance modulus. A
similar diagram for the BP scenario (middle panel) reveals that a
background/foreground star transited by a planet can produce a signal
mimicking the one observed for a wide range of masses between 0.2 and
1.4~$M_{\sun}$, provided it is within 5.4 magnitudes of the brightness
of the target. In this case the observational constraints from
spectroscopy and color remove large portions of parameter space.
Finally, the right panel of Figure\;\ref{fig:blender} shows that the
allowed area of parameter space for the HTP configurations is a narrow
swath in which the companion mass is larger than about 0.37~$M_{\sun}$
and the planet size is between 0.2 and 0.65~$R_{\rm Jup}$
(2.2--7.3~$R_{\earth}$). However, much of this region is excluded
either by the spectroscopic constraint or the color constraint.

Cross-sections of parameter space for the other KOIs are similar to
those shown for KOI-0172.02, though we note that in one case
(KOI-0463.01) \blender\ indicates that {\it all\/} blends involving
background/foreground eclipsing binaries (BEB) or stars transited by a
planet (HTP, BP) can be ruled out simply from the shape of the
transit.

\setlength{\tabcolsep}{-4pt}

\setlength{\tabcolsep}{6pt}

The blend frequencies obtained for each KOI and each type of false
positive scenario are presented in Table\;\ref{tab:validations}.  The
BEB quantities include the contribution from blended eclipsing
binaries with equal primary and secondary eclipse depths and twice the
orbital period, which is another potential source of confusion.  As we
indicated earlier, the overall blend frequency is seen to be dominated
by the HTP scenarios (planet orbiting a physical companion to the
target), which are typically two to three orders of magnitude 
more common than the other false positive scenarios taken
together. Also listed in the table are the expected planet priors
(``PL''), the odds ratios defined as ${\rm PL}/({\rm HTP}+{\rm
  BP}+{\rm BEB})$, and the statistical significance of the
validations.

As mentioned above, for a 3$\sigma$ validation (99.73\% confidence
level) we require a planet frequency that is at least $1/(1/99.73\%-1)
\approx 370$ times larger than the total blend frequency. All except
for three of our 18 targets meet this threshold. KOI-4054.01, 4450.01,
and 5236.01 are validated to somewhat lower confidence levels, though
still above 99.4\%. 
Two of the candidates, KOI-2626.01 and KOI-4550.01, are among those
newly validated at the 99.73\% confidence level or greater, and are now
given planet designations Kepler-1652\,b and Kepler-1653\,b,
respectively (see Table\;\ref{tab:targets}).
For the remainder of the paper we continue to
refer to all validated planets by their KOI designations.

\begin{deluxetable*}{lcccccccc}
\tablecaption{Blend Frequencies, Planet Priors, Odds Ratios and
  Significance Level of the Validations.\label{tab:validations}}
\tablehead{
\colhead{Candidate} &
\colhead{HTP} &
\colhead{BP} &
\colhead{BEB} &
\colhead{PL} &
\colhead{Odds Ratio} &
\colhead{Significance} &
\colhead{PL$_{\rm comp}$} &
\colhead{$\mathcal{P}[{\rm targ}]$}
}
\startdata
KOI-0172.02  &  2.94E$-$07 &  5.71E$-$11 &   \nodata   &  2.27E$-$04 &    772 &  99.87\%  &  \nodata     &  \nodata  \\
KOI-0438.02  &  6.32E$-$08 &  9.74E$-$12 &  5.88E$-$12 &  7.34E$-$05 &   1161 &  99.91\%  &  \nodata     &  \nodata  \\
KOI-0463.01  &   \nodata   &   \nodata   &   \nodata   &  1.43E$-$05 &\nodata & 100.00\%  &  \nodata     &  \nodata  \\
KOI-0812.03  &  1.65E$-$07 &  7.30E$-$12 &   \nodata   &  2.31E$-$04 &   1400 &  99.93\%  &  \nodata     &  \nodata  \\
KOI-0854.01  &  1.03E$-$07 &   \nodata   &   \nodata   &  4.06E$-$05 &    394 &  99.75\%  &  6.32E$-$05  &   39.1\%  \\
KOI-2418.01  &   \nodata   &  1.41E$-$11 &   \nodata   &  2.05E$-$04 &\nodata & 100.00\%  &  4.63E$-$06  &   97.8\%  \\
KOI-2626.01  &  9.89E$-$07 &  8.08E$-$10 &  1.83E$-$09 &  4.84E$-$04 &    488 &  99.80\%  &  3.94E$-$04  &   55.2\%  \\
KOI-2650.01  &  6.01E$-$07 &  9.97E$-$10 &  1.20E$-$10 &  7.38E$-$04 &   1226 &  99.92\%  &  \nodata     &  \nodata  \\
KOI-3010.01  &  1.94E$-$07 &  4.74E$-$10 &  1.06E$-$10 &  4.10E$-$04 &   2107 &  99.95\%  &  2.75E$-$04  &   59.9\%  \\
KOI-3282.01  &  1.58E$-$06 &  4.37E$-$10 &  1.05E$-$10 &  8.15E$-$04 &    516 &  99.81\%  &  \nodata     &  \nodata  \\
KOI-3497.01  &   \nodata   &  3.48E$-$08 &   \nodata   &  8.54E$-$04 &  24540 & 100.00\%  &  5.18E$-$04  &   62.2\%  \\
KOI-4036.01  &  3.93E$-$07 &  3.40E$-$10 &   \nodata   &  5.36E$-$04 &   1363 &  99.93\%  &  \nodata     &  \nodata  \\
KOI-4054.01  &  3.44E$-$07 &  7.15E$-$11 &   \nodata   &  9.54E$-$05 &    277 &  99.64\%  &  \nodata     &  \nodata  \\
KOI-4356.01  &  1.97E$-$07 &  5.84E$-$10 &  3.01E$-$12 &  1.90E$-$04 &    962 &  99.90\%  &  \nodata     &  \nodata  \\
KOI-4450.01  &  3.06E$-$06 &  8.16E$-$09 &  1.22E$-$12 &  5.69E$-$04 &    185 &  99.46\%  &  \nodata     &  \nodata  \\
KOI-4550.01  &  5.07E$-$07 &  2.11E$-$09 &  1.80E$-$10 &  9.18E$-$04 &   1803 &  99.94\%  &  8.29E$-$04  &   52.5\%  \\
KOI-5236.01  &  5.89E$-$07 &  1.56E$-$10 &   \nodata   &  1.55E$-$04 &    263 &  99.62\%  &  \nodata     &  \nodata  \\
KOI-5856.01  &  8.74E$-$08 &  6.41E$-$11 &   \nodata   &  3.16E$-$04 &   3613 &  99.97\%  &  \nodata     &  \nodata   
\enddata

\tablecomments{For several of the candidates the blend frequencies for
  the HTP, BP, and/or BEB scenarios are vanishingly small and are not
  reported. In eight cases in which the targets have a close companion
  detected from high-resolution imaging we list the estimated planet
  frequency PL$_{\rm comp}$ assuming the planet transits the
  companion. For KOI-2626, which has two companions that can
  potentially mimic the transit light curve if transited by a planet,
  we have added the corresponding frequencies.  $\mathcal{P}[{\rm
      targ}]$ is the probability that the planet transits the target
  instead of a companion.}

\end{deluxetable*}

The validation results presented above are based on a comparison
between the likelihood that the transit signals originate in the
targets and the likelihood that they come from an {\it unseen}
eclipsing object blended with the target. However, eight of our stars
have one or more {\it known} close companions discovered with
high-resolution imaging that cannot be excluded as the source of the
signals by the centroid motion analysis described in
Section~\ref{sec:centroids}.  If the transit signals come from planets
orbiting these companions, the size of the planet could be larger than
it appears by a factor that depends on the physical size and
brightness of the companion star.

Studies by \cite{Horch:2014} and \cite{Hirsch:2017} have concluded
that the vast majority of sub-arc second companions to KOIs are bound
to their primary stars. All but one of the companions to our targets
are within 1\arcsec, the exception being the $\rho \approx 1\farcs94$
neighbor of KOI-5236, which is also the faintest of the ones not ruled
out by the centroid motion analysis ($\Delta K \approx 6.4$~mag).  For
the seven companions that are within 0\farcs43 (see
Table\;\ref{tab:closecompanions}) the conclusion that they are bound
is further supported by the fact that we see no signs of these stars
in the Keck/HIRES spectra even though they are all bright enough to
have been discovered. This implies that their radial velocities must
be similar to those of the targets, which argues for physical
association.  Under the assumption that companions within
$\sim$1\arcsec\ are bound, examination of the \blender\ constraints
for the HTP scenario indicates that the $\rho \approx 0\farcs15$
neighbor of KOI-0854 can be excluded as a potential false positive
because no planet transiting it can mimic the \kepler\ light curve
well enough. The second companion to this KOI remains viable. For the
$\rho \approx 1\farcs94$ companion to KOI-5236 the situation is less
clear, as \cite{Hirsch:2017} found that wider companions out to
2\arcsec\ are about equally likely to be bound as unbound. However,
the \blender\ constraints for both the HTP and the BP scenarios
indicate that this star is too faint to be the source of the transit
for any reasonable size planet. Thus, we are left with a total of
seven KOIs in which we cannot tell a priori whether the planet is
around the target or one of the neighboring stars within about
1\arcsec.

To estimate the chance that the planet transits the KOI rather than a
companion, we inferred the properties of the companions (size,
brightness) using the same isochrone appropriate for the primary star
and computed the planet prior in the same way as done previously for
the target, but this time assuming the planet is around the companion.
The amount of dilution of the signal was adjusted accordingly to
account for the larger contamination now coming from a brighter star
in the aperture (the KOI). These planet frequencies are listed in
Table\;\ref{tab:validations} under the column labeled PL$_{\rm
  comp}$. For the case of KOI-2626 with two close companions we have
combined the companion planet frequencies. The chance that the planet
is around the KOI as opposed to one of the companions was computed as
$\mathcal{P}[{\rm targ}] = {\rm PL}/({\rm PL} + {\rm PL}_{\rm
  comp})$. In the case of KOI-2418.01 the statistics strongly suggest
the planet indeed orbits the KOI; for the other six cases we consider
the results to be too close to 50\% to be conclusive. The possibility
remains, therefore, that the planets in these systems are larger than
they appear, and we quantify this and discuss the implications in
the next section.

\section{Light-curve fits and planetary parameters}
\label{sec:fits}

The \kepler\ light curves of the KOIs, detrended as described in
Section\;\ref{sec:photometry}, were modeled following the procedure
detailed in Sect.\;4 of \citet{Rowe:2014} that uses the analytic
quadratic limb-darkening transit prescription of \citet{Mandel:2002}.
The technique is capable of handling systems with multiple transiting
planets fitting them simultaneously and parametrizing the transit
model with the mean stellar density $\rho_{\star}$, on the assumption
that all planets in a given system transit the same star and that
their total mass is much smaller than that of the host star. For the
three KOIs in our sample hosting multiple planets (KOI-0172, KOI-0438,
KOI-0812) we have further assumed non-interacting Keplerian orbits.
Our model uses the quadratic limb-darkening reparametrization of
\cite{Kipping:2013} with coefficients $q_1$ and $q_2$ to characterize
the brightness profile of the host star.  For each transiting planet
$n$ in a multi-planet system we solved for the mean stellar density
$\rho_{\star}$ (subject to a prior given by our spectroscopic work),
the time of the first observed transit $T_{0,n}$, the orbital period
$P_n$, a photometric zero point, the scaled planetary radius
$(R_p/R_{\star})_n$, the impact parameter $b_n$, and the shape
parameters ($e_n$, $\omega_n$) parametrized as $e_n \sin\omega_n$ and
$e_n \cos\omega_n$.

To explore the impact of limb-darkening and eccentricity on our
measured model parameters, particularly on $R_p/R_{\star}$, we
computed three sets of models:
\begin{enumerate}[leftmargin=15pt,itemsep=-1pt]
	\item Orbit assumed to be circular, and limb-darkening
          coefficients held fixed based on the stellar properties
          listed in Table\;\ref{tab:physical} and the limb-darkening
          tables of \cite{Claret:2011} for the \kepler\ bandpass;
	\item Orbit assumed to be circular, and limb-darkening
          coefficients solved for;
	\item Both the eccentricity and the limb-darkening
          coefficients considered as adjustable parameters.
\end{enumerate}
For all three models we used a prior on $\rho_{\star}$ based on the
stellar properties presented in Table \ref{tab:physical}.

\begin{figure*}
\epsscale{1.15}
\plotone{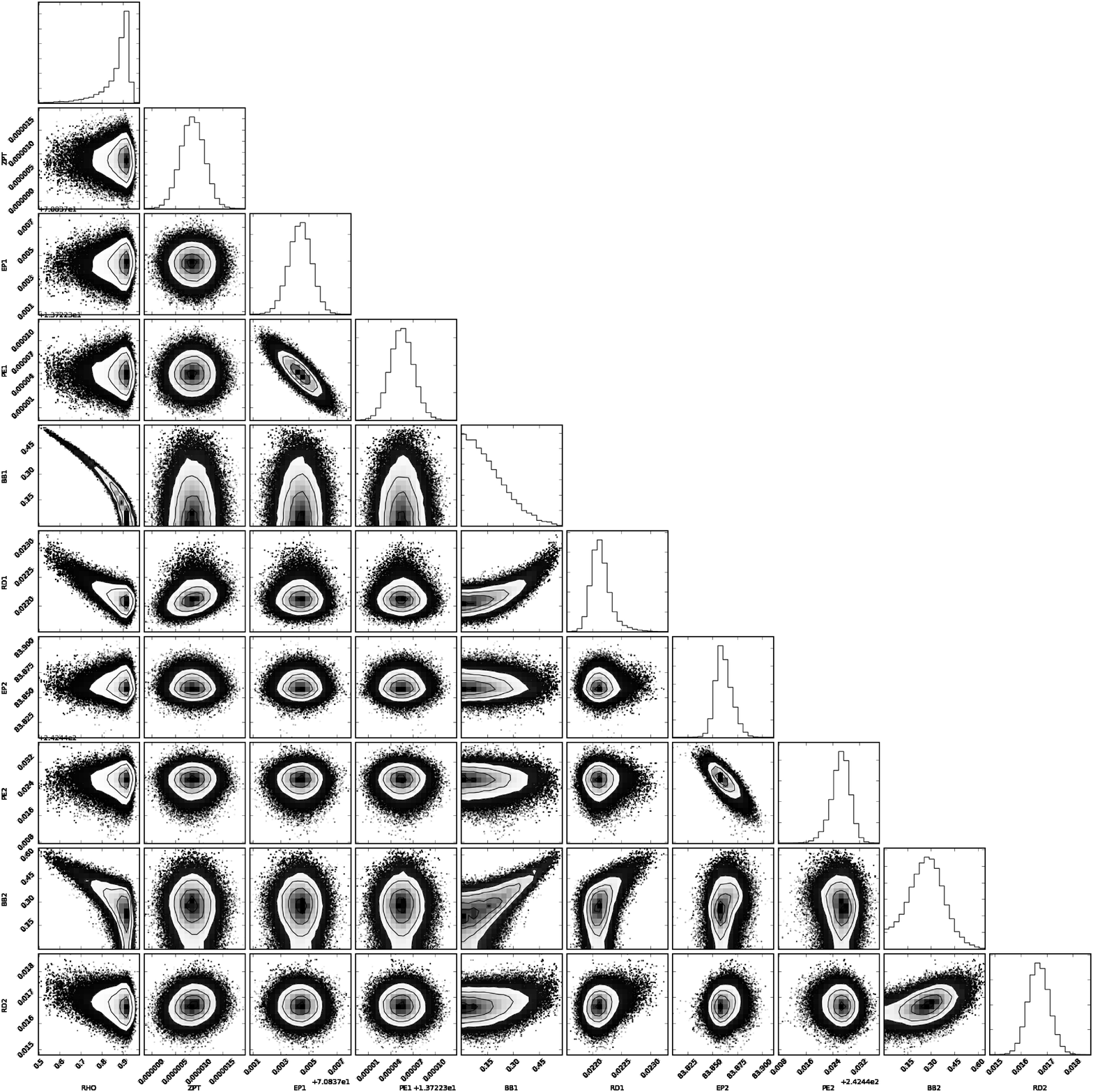} 

\figcaption[]{``Corner plot''
  \citep{Foreman-Mackey:2016}\footnote{\url
    https://github.com/dfm/corner.py} for the KOI-0172 multiplanet
  model fit illustrating the correlations among the fitted variables,
  for the case of circular orbits and fixed limb-darkening
  coefficients.  Histogram panels represent the posterior distribution
  for each parameter. RHO is the mean stellar density
  ($\rho_{\star}$), ZPT is the photometric zero point, and EPX is the
  time of transit center, with X designating the modeled planet (1 or
  2).  PEX, BBX and RDX are the orbital period (days), impact
  parameter ($b$), and scaled planetary radius ($R_p/R_{\star}$),
  respectively. Clear correlations among various parameters can be
  seen, such as between $\rho_{\star}$ and $b$.  The correlation
  between $b$ for the two planets in the KOI-0172 system shows that
  the relative inclinations are better measured than absolute
  values. \label{fig:corner}}

\end{figure*}

We obtained best fit model parameters in each case using the code {\tt
  TRANSITFIT5} \citep{Rowe:2016}, which implements a
Levenberg-Marquardt $\chi^2$ minimization routine \citep{More:1980}.
Posterior distributions of the model parameters were derived from a
Markov Chain Monte Carlo (MCMC) analysis.  The MCMC algorithm is
similar to the approach described by \citet{Ford:2005} and
\citet{Rowe:2014}.  The program {\tt TRANSITMCMC5} \citep{Rowe:2016}
was used to generate two Markov chains with lengths of $10^6$.  The
first 10\% of each chain was discarded as burn-in.  A visual
examination of the chains showed good mixture, and the Gelman-Rubin
convergence criterion yielded $R_c$ values of 1.01 or lower for all
fitted parameters.  Figure\;\ref{fig:corner} displays histograms of
the chain values for each parameter, and scatter plots to unveil
potential correlations for the model parameter distributions in
KOI-0172.02, for the case of $e = 0$ and fixed limb-darkening.
KOI-0172 has two known transiting planets.  Figure\;\ref{fig:corner}
shows a clear correlation between $\rho_{\star}$, $b_1$, and $b_2$,
which is expected as the transit duration is a geometrical measure of
$\rho_{\star}$ for a circular orbit.  There is also a correlation
between $b_1$ and $b_2$, which demonstrates that relative inclinations
are better measured than absolute inclinations.  The correlation
between $R_p/R_{\star}$ and $b_1$, and similarly with $b_2$, is due to
stellar limb-darkening.

\begin{figure}
\epsscale{1.10}
\plotone{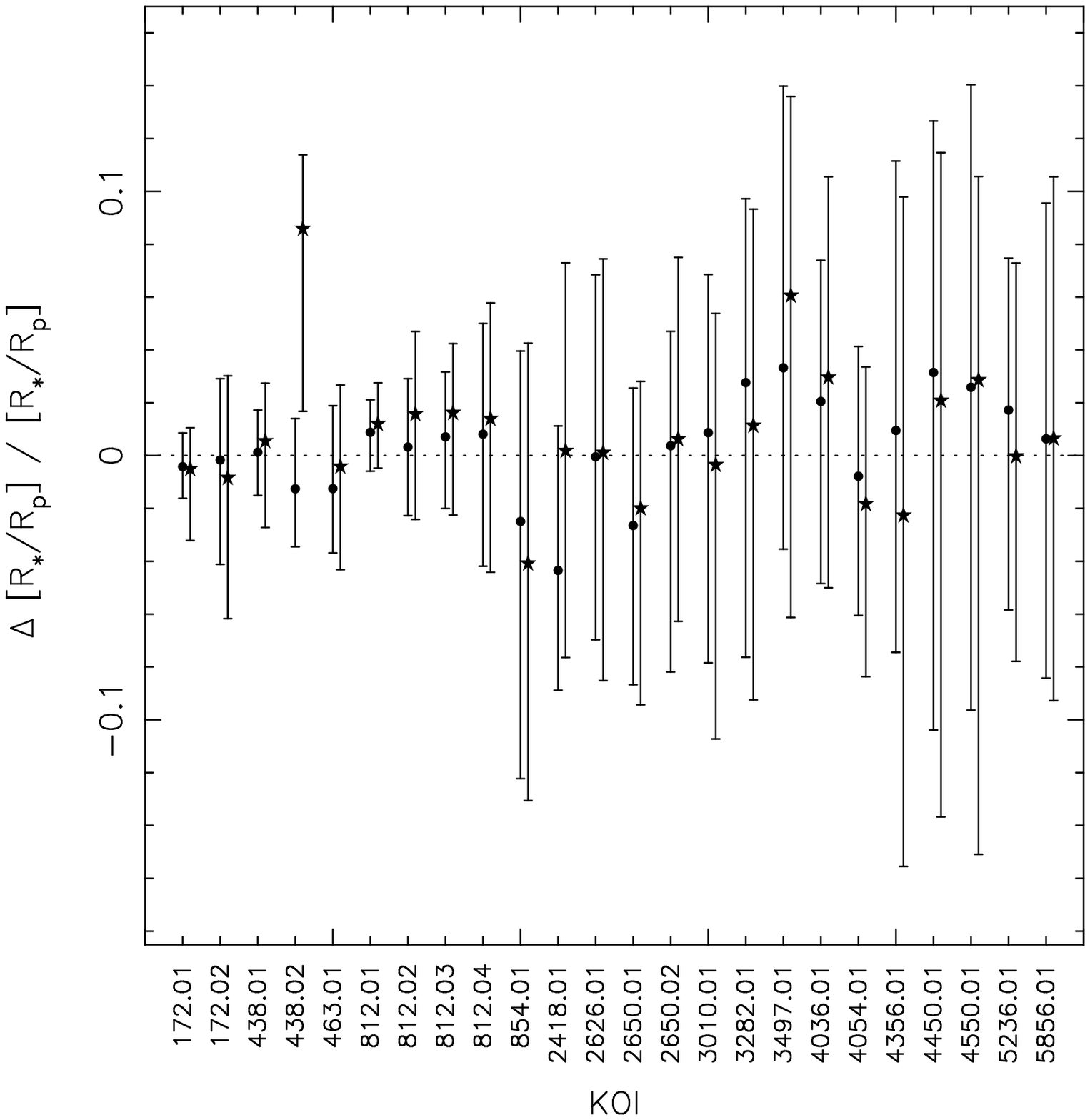}

\figcaption[]{Fractional $R_p/R_{\star}$ differences and 1$\sigma$
  uncertainties between three alternate models fits for each KOI.
  Dots represent the difference between a model with fitted quadratic
  limb-darkening (Model\;2) and one with fixed quadratic
  limb-darkening parameters (Model\;1), both with the eccentricity
  held at zero.  Star symbols represent the difference between a model
  with the eccentricity and quadratic limb-darkening coefficients left
  free (Model\;3) and a model with a circular orbit and fixed
  quadratic limb-darkening parameters (Model\;1). In general the
  uncertainty in $R_p/R_{\star}$ increases with model complexity, but
  in all cases the uncertainty in the stellar radius dominates the
  error budget.  Most differences in $R_p/R_{\star}$ are consistent
  with zero. The largest excursion is seen for KOI-438.02, which
  weakly (less than 2$\sigma$ confidence level) suggests either the
  detection of a slight eccentricity or a bias in our adopted
  $\rho_{\star}$ used as a prior.\label{fig:comparemodels}}
\end{figure}

Models\;2 and 3 above allow us to explore how well we can measure
eccentricity and limb-darkening, and to understand potential biases in
the measured planetary radii.  Figure \ref{fig:comparemodels} shows
the relative change in the measured value of $R_p/R_{\star}$ when
comparing Model\;2 against Model\;1 (dots) and Model\;3 versus
Model\;1 (stars) for each modeled planet.  None of the model results
differ by more than 1$\sigma$, except for KOI-438.02.  In general, the
uncertainty in $R_p/R_{\star}$ increases with model complexity. The
fractional change in $R_p/R_{\star}$ between the three models is less
than 5\%, with most models showing differences smaller than 1.5\%.
Thus, the dominant source of error in the measurement of the planetary
radius ($R_p$) is the error in the stellar radius.

The results of our transit model fits are presented in
Table\;\ref{tab:planetproperties1}, in which for completeness we
include the additional members of the four multi-planet systems as
well (KOI-0172, 0438, 0812, and 2650). The values reported in the table
correspond to the mode of the posterior distribution for each
parameter based on our most simplistic model (Model\;1) with a
circular orbit and fixed quadratic limb-darkening coefficients. If
future studies significantly increase the fidelity of the stellar
parameters (e.g., through improvements coming from {\it Gaia}
parallaxes), then a more complex photometric model may be warranted.
In the case of KOI-438.02 Model\;3 points toward a systematic increase
in the measured scaled planetary radius, which is due to this model
having a marginal preference for a non-circular orbit.  This may
indicate either a real eccentricity, or an error in the stellar
parameters, for the following reason.  The transit duration is
proportional to the mean stellar density, and because we have applied
a prior on $\rho_{\star}$ informed by our spectroscopic work, any
error in this quantity could lead to a mismatch in the expected
transit duration for a circular orbit.  The model accounts for the
difference by making the orbit eccentric.  Breaking this degeneracy
requires either a more accurate measure of the fundamental stellar
parameters or an independent determination of the eccentricity of the
orbit.

\input planettable1b.tex

\input planettable2b.tex

\begin{figure*}
\epsscale{1.15}
\plotone{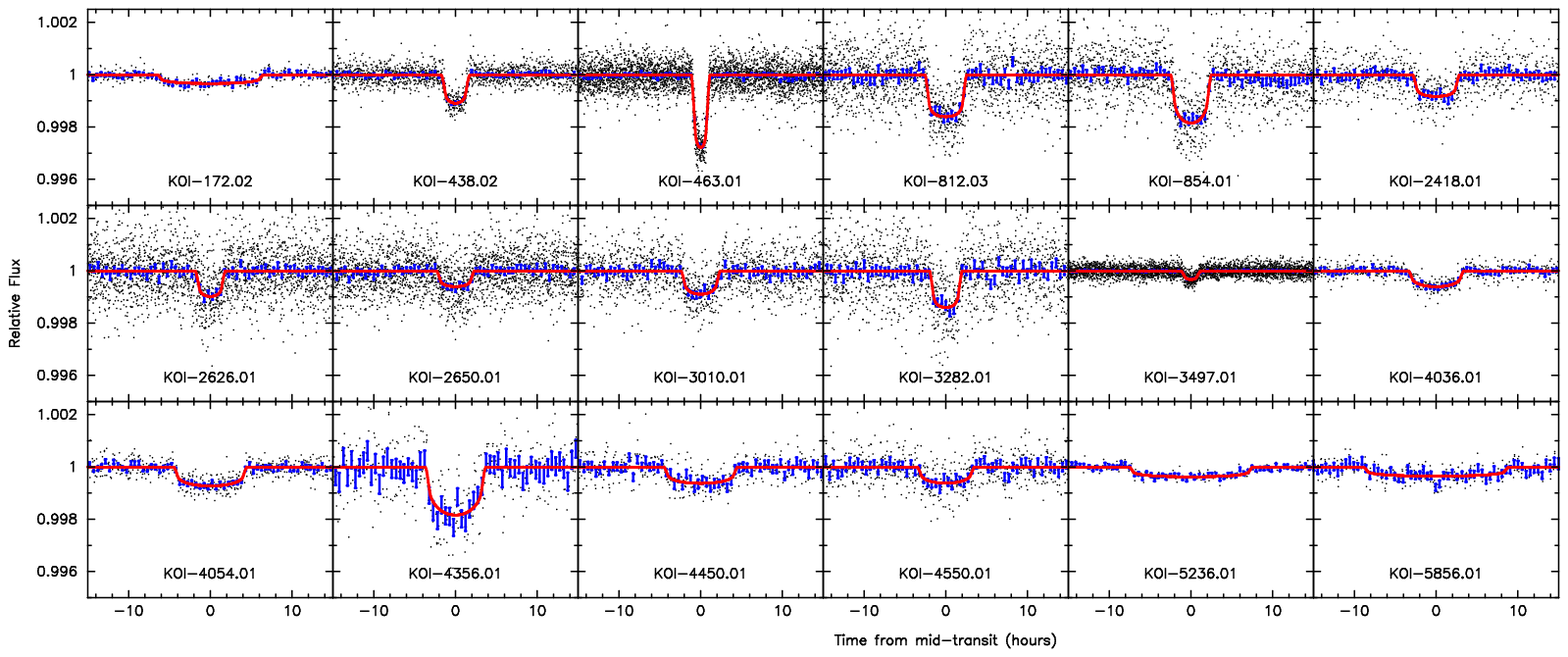} 

\figcaption[]{Light-curve fits (red lines) and \kepler\ observations
  within 15 hours of the center-of-transit time for our 18 KOIs. The
  observations shown as black dots are a combination of short- and
  long-cadence data. Blue symbols represent data averaged in 30\;min
  bins, with 1$\sigma$ error bars computed from the standard
  deviation.\label{fig:allplanets}}

\end{figure*}

Table\;\ref{tab:planetproperties2} presents additional derived
parameters from our fit including the scaled semimajor axis
$a/R_{\star}$, the fitted transit depth and total duration
$\delta_{\rm tot}$, the duration of ingress/egress $\delta_{12}$
(interval between first and second contacts), the semimajor axis $a$
and inclination angle $i$, the incident flux $S_{\rm eff}$ in units of
the Earth's, and the equilibrium temperature of the planet $T_{\rm
  eq}$. The uncertainty in the latter accounts for a possible range of
albedos between 0 and 0.5, as well as energy redistribution factors
ranging from full redistribution to the planet being tidally locked.
Our light-curve fits are illustrated in Figure\;\ref{fig:allplanets}.
Because the modal values reported in
Table\;\ref{tab:planetproperties1} will generally not result in a best
fit light curve, for the generation of Figure\;\ref{fig:allplanets} we
have chosen to plot a transit model that is based on best fit
parameters.

The planetary radii listed in Table\;\ref{tab:planetproperties1}
assume that in each case the planet orbits the KOI. If instead the
planet orbits a nearby companion, the true planetary size can be
considerably larger because of the generally greater dilution of the
signal caused by the brighter target star. The correction factor
depends on the relative brightness $\Delta K\!p$ between the close
neighbor and the target and on the size of the companion \citep[see
  also][]{Ciardi:2015, Furlan:2017}.  Table\;\ref{tab:dilution} gives
the relative brightness and angular separation of the close companions
that are not excluded by the centroid motion analysis or by \blender,
along with the radius correction factors $X_R$ and the $R_p$ values
that apply if the planet orbits the companion. These planetary radii
are seen to be typically $\sim$50\% larger than those in
Table\;\ref{tab:planetproperties1}. In most cases these larger sizes
would compromise potential habitability because they would imply the
planets are more likely gaseous than rocky.

\setlength{\tabcolsep}{3pt}

\begin{deluxetable}{lccccccc}
\tablewidth{0pc}

\tablecaption{Planetary Radii Corrected for Dilution, for Stars in
  Which the Planet Transits a Fainter Nearby Companion Rather than the
  Target.\label{tab:dilution}}

\tablehead{
\colhead{} &
\colhead{$\rho$} &
\colhead{$\Delta K\!p$} &
\colhead{$M_{\rm comp}$} &
\colhead{$R_{\rm comp}$} &
\colhead{} &
\colhead{$R_p$ comp.}
\\
\colhead{Name} &
\colhead{(\arcsec)} &
\colhead{(mag)} &
\colhead{($M_{\sun}$)} &
\colhead{($R_{\sun}$)} &
\colhead{$X_R$} &
\colhead{($R_{\earth}$)} 
}
\startdata
KOI-0854.01  &  0.016  &  0.39  &  0.49  &  0.46  &  1.44  &  3.96 \\
KOI-2418.01  &  0.108  &  3.32  &  0.12  &  0.14  &  1.48  &  1.98 \\
KOI-2626.01  &  0.164  &  1.46  &  0.24  &  0.24  &  1.66  &  2.66 \\
KOI-2626.01  &  0.206  &  0.62  &  0.33  &  0.32  &  1.49  &  2.38 \\
KOI-3010.01  &  0.334  &  0.43  &  0.55  &  0.52  &  1.46  &  3.14 \\
KOI-3497.01  &  0.795  &  1.31  &  0.25  &  0.25  &  1.35  &  1.08 \\
KOI-4550.01  &  1.045  &  0.89  &  0.63  &  0.61  &  1.61  &  3.48  
\enddata

\tablecomments{The columns list the angular separation of the
  companion and the estimated brightness difference relative to the
  target in the $K\!p$ band, the estimated mass and radius of the
    companion star, the correction factor $X_R$ for the planetary
  radius, and the corrected radius if the planet transits the
  companion.}

\end{deluxetable}

\setlength{\tabcolsep}{6pt}

\section{Habitability}
\label{sec:habitability}

The MCMC analysis of the \kepler\ light curves together with the
stellar properties described earlier provide probability distributions
for the planetary and stellar parameters.  Here we have combined these
distributions with calculations for the Habitable Zone in order to
quantify the probability that the planets lie in the HZ, and we assume
for these estimates that the planets orbit the target.  The boundaries
of the HZ were calculated using the empirically derived relations
presented by \citet{Kopparapu:2013b, Kopparapu:2014}. We used both the
``conservative'' (CHZ) and ``optimistic'' (OHZ) boundaries for the HZ
detailed by \citet{Kane:2016}, whose locations are determined based
upon assumptions regarding how long Venus and Mars may have been able
to retain liquid water on their surfaces. The stellar fluxes, $S_{\rm
  eff}$, at each HZ boundary were determined from a polynomial
relationship involving the effective temperature $T_{\rm eff}$. The HZ
boundaries were then calculated from
\begin{equation}
  d = (L_{\star}/S_{\rm eff})^{1/2} \ \mathrm{au}
\end{equation}
where $L_{\star}$ is the luminosity of the star in solar units.

\begin{figure*}
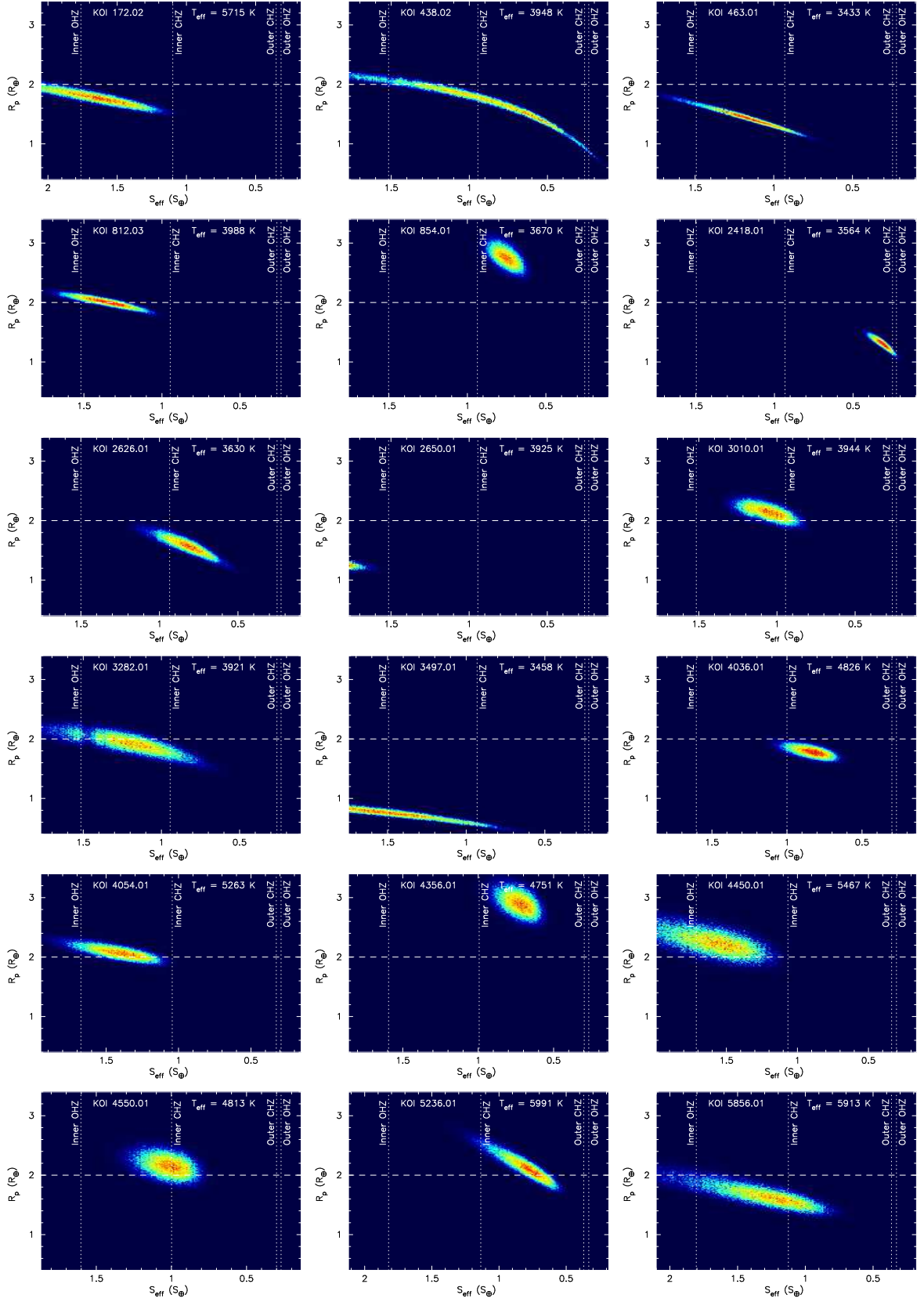

\centering
\begin{tabular}{ccc}

\includegraphics[width=3.5cm,angle=-90]{plot_KOI_172_02.ps} &
\includegraphics[width=3.5cm,angle=-90]{plot_KOI_438_02.ps} &
\includegraphics[width=3.5cm,angle=-90]{plot_KOI_463_01.ps} \\

\includegraphics[width=3.5cm,angle=-90]{plot_KOI_812_03.ps} &
\includegraphics[width=3.5cm,angle=-90]{plot_KOI_854_01.ps} &
\includegraphics[width=3.5cm,angle=-90]{plot_KOI_2418_01.new.ps} \\

\includegraphics[width=3.5cm,angle=-90]{plot_KOI_2626_01.ps} &
\includegraphics[width=3.5cm,angle=-90]{plot_KOI_2650_01.ps} &
\includegraphics[width=3.5cm,angle=-90]{plot_KOI_3010_01.ps} \\

\includegraphics[width=3.5cm,angle=-90]{plot_KOI_3282_01.ps} &
\includegraphics[width=3.5cm,angle=-90]{plot_KOI_3497_01.ps} &
\includegraphics[width=3.5cm,angle=-90]{plot_KOI_4036_01.ps} \\

\includegraphics[width=3.5cm,angle=-90]{plot_KOI_4054_01.ps} &
\includegraphics[width=3.5cm,angle=-90]{plot_KOI_4356_01.ps} &
\includegraphics[width=3.5cm,angle=-90]{plot_KOI_4450_01.ps} \\

\includegraphics[width=3.5cm,angle=-90]{plot_KOI_4550_01.ps} &
\includegraphics[width=3.5cm,angle=-90]{plot_KOI_5236_01.ps} &
\includegraphics[width=3.5cm,angle=-90]{plot_KOI_5856_01.new.ps} \\

\end{tabular}

\figcaption[]{Heat maps of the joint posterior distributions for the
  planetary radii $R_p$ and stellar fluxes $S_{\rm eff}$ for each of
  our KOIs.  They are shown in units of Earth radii and the incident
  flux received by the Earth from the Sun, respectively. The vertical
  dotted lines show the locations of the conservative HZ (CHZ) and
  optimistic HZ (OHZ) boundaries, and the horizontal dashed line marks
  a planetary radius of $2\;R_{\earth}$, for
  reference. \label{fig:habitability}}

\end{figure*}

Shown in Figure\;\ref{fig:habitability} are representations of the
joint posterior distributions for the planetary radius and the
incident flux for each of the 18 KOIs analyzed in this work. In each
panel we used the same range of planetary radii on the vertical axes,
and the horizontal axes are scaled so that the OHZ boundaries are
aligned to facilitate the comparison.  The inner and outer CHZ and OHZ
boundaries are indicated by vertical dotted lines, and a horizontal
dashed line shows the location where $R_p = 2.0\;R_{\earth}$, for
reference. The methodology of \citet{Kane:2016} defines a category of
HZ candidates that lie inside the OHZ and have radii less than
$2\;R_{\earth}$. The rationale behind this radius boundary is that in
many cases (depending on composition, incident flux, etc.) the
transition between a rocky and a gaseous planet occurs somewhere
between 1.5--2.0\;$R_{\earth}$ \citep{Weiss:2014, Rogers:2015,
  Wolfgang:2015, Fulton:2017}. Given the uncertainties on the
planetary and stellar properties, \citet{Kane:2016} chose an
upper radius limit of $2\;R_{\earth}$ to define their list of
likely terrestrial candidates within the OHZ (see their Table\;2).

\setlength{\tabcolsep}{3pt}
\setlength{\tabcolsep}{6pt}

Using a procedure similar to that of \citet{Torres:2015} we further
calculated the probability that each candidate falls within the region
defined by the OHZ and the $R_p \leq 2.0\,R_{\earth}$ boundaries. This
was done for each type of boundary by determining the number of
posterior realizations that lie within the corresponding region. The
results of these calculations are presented in
Table\;\ref{tab:habitability}. They show that 15 of the 18 candidates
have a greater than 50\% chance of lying within the OHZ, 10 of the
candidates meet the $R_p \leq 2.0\,R_{\earth}$ criterion, and 7 meet
both criteria.

\begin{deluxetable}{lcc}
\tablewidth{0pc}

\tablecaption{Probability of the KOIs meeting the OHZ and $R_p \leq
  2\,R_{\earth}$ Criteria.\label{tab:habitability} }

\tablehead{
\colhead{Name} &
\colhead{Within OHZ (\%)} &
\colhead{$R_p \leq 2 R_{\earth}$ (\%)}
}
\startdata
KOI-0172.02  &  \phn46.1  &  \phn74.5 \\
KOI-0438.02  &  \phn73.2  &  \phn65.4 \\
KOI-0463.01  &  \phn85.0  &  \phn99.8 \\
KOI-0812.03  &  \phn79.5  &  \phn46.0 \\
KOI-0854.01  & 100.0  &   \phn\phn0.0 \\
KOI-2418.01  &  \phn97.4  & 100.0 \\
KOI-2626.01  & 100.0  &  \phn98.7 \\
KOI-2650.01  &   \phn\phn0.1  & 100.0 \\
KOI-3010.01  & 100.0  &  \phn14.1 \\
KOI-3282.01  &  \phn79.9  &  \phn62.4 \\
KOI-3497.01  &  \phn38.7  & 100.0 \\
KOI-4036.01  & 100.0  &  \phn97.8 \\
KOI-4054.01  &  \phn83.6  &  \phn22.9 \\
KOI-4356.01  & 100.0  &   \phn\phn0.0 \\
KOI-4450.01  &  \phn53.7  &   \phn\phn9.2 \\
KOI-4550.01  &  \phn99.9  &  \phn19.2 \\
KOI-5236.01  &  \phn99.9  &  \phn23.3 \\
KOI-5856.01  &  \phn72.7  &   \phn79.3  
\enddata
\end{deluxetable}

\section{Discussion}
\label{sec:discussion}

Of the 18 planet candidates on our target list we have validated 15 at
a confidence level of 99.73\% or higher (3$\sigma$). The other three
(KOI-4054.01, KOI-4450.01, and KOI-5236.01) are currently validated at
a slightly lower levels of 99.5--99.6\% (2.8$\sigma$--2.9$\sigma$),
which could be improved with further spectroscopic or deep imaging
observations to help rule out additional false positives. The 18
objects span a range of planetary radii from about 0.80 to
2.9~$R_{\earth}$.  Seven of them (KOI-0438.02, 0463.01, 2418.01,
2626.01, 3282.01, 4036.01, and 5856.01) have the best chance ($>
50\%$) of being in the optimistic HZ and at the same time being
smaller than 2~$R_{\earth}$. As such they are valuable for studies of
the rate of occurrence of small planets in the habitable zone of their
host stars, one of the primary science goals of the
\kepler\ Mission. All but one of these six orbit an M dwarf, and their
orbital periods range from 18 to 169 days. Among them, we note that
KOI-0438.02 was not included in the catalog of small HZ planets by
\cite{Kane:2016} because the stellar radius and pipeline transit
modeling parameters adopted in that work
\citep[DR25;][]{Thompson:2017} placed the planet just above the
2\;$R_{\earth}$ cutoff they used, whereas we now measure a size just
below it.

\begin{figure}
\epsscale{1.15}
\plotone{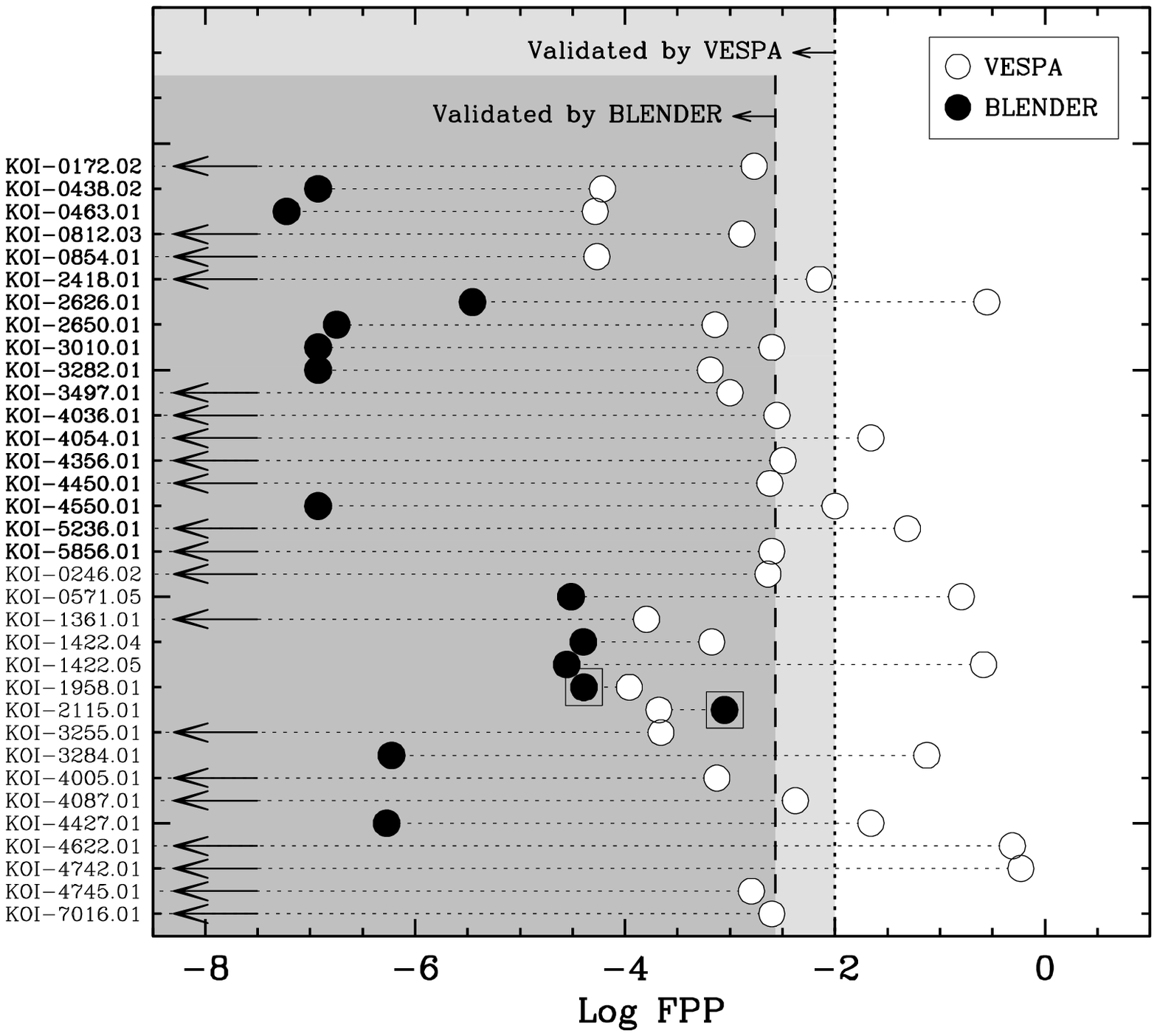}
                                             
\figcaption[]{Comparison between false positive probabilities from
  \blender\ (excluding HTP and BP scenarios) and \vespa\ for 34 KOIs
  subjected to validation by both methods, as labeled on the left.
  Those from the present work are indicated in bold at the top. Arrows
  represent upper limits from \blender, and the two values marked with
  squares \citep[KOI-1958.01 and KOI-2115.01;][]{Meibom:2013}
  correspond to special cases in which the \blender\ validations used
  more conservative assumptions than usual for the planet priors (see
  text). For candidates to be declared validated with \vespa\ they are
  required to be to the left of the dotted line (${\rm FPP} \leq
  0.01$; light gray area), whereas validated planets from
  \blender\ are those to the left of the dashed line (${\rm FPP} \leq
  0.0027$; darker gray area). \label{fig:fpp}}

\end{figure}

The 18 KOIs studied in this work with \blender\ have been previously
subjected to a similar statistical validation exercise using the
\vespa\ procedure \citep{Morton:2016}\footnote{\vespa\ is an
  open-source code available to the community; the more
  computer-intensive \blender\ code was custom-designed to run on the
  Pleiades supercomputer at the NASA Ames Research Center (CA).}, with
the result that all but three of them (KOI-2626.01, KOI-4054.01, and
KOI-5236.01) were considered validated with that algorithm at the 99\%
or higher confidence level. Given that the two methodologies are
completely independent, it is of interest to compare the false
positive probabilities (FPPs) returned by each approach. As pointed
out earlier, one of the key conceptual differences between
\blender\ and \vespa\ is that \vespa\ does not consider as false
positives configurations that involve a planet transiting a star
different than the target along the same line of sight. These
correspond to the scenarios we have referred to here as BP and HTP,
and we have argued that for applications such as ours were the size of
the planet is important they should be included in calculating the
FPP.  To place the FPPs from both methods on the same footing for this
comparison, we have recomputed those from \blender\ as ${\rm FPP} =
{\rm BEB}/({\rm PL}+{\rm BEB})$, leaving out the blend frequencies
from the HTP and BP scenarios.  The top portion of
Figure\;\ref{fig:fpp} shows the FPPs from \blender\ and \vespa\ for
our 18 targets (names in boldface). An additional 16 candidates from
the literature have been examined in recent years both with
\vespa\ and with a version of \blender\ essentially the same as that
used here \citep{Barclay:2013, Meibom:2013, Torres:2015,
  Jenkins:2015}, and we include them in the comparison as well.
Arrows indicate upper limits from \blender\ for cases in which blends
with background/foreground eclipsing binaries are virtually all
excluded simply from the shape of the transits.

Overall the more detailed treatment given to each candidate in
\blender, and perhaps the more elaborate assumptions, result in
validations that are typically stronger than those from \vespa, with
FPP values that are an order of magnitude or more
smaller.\footnote{There are two exceptions marked in
  Figure\;\ref{fig:fpp} with squares, for which the \blender\ and
  \vespa\ FPP values are most similar. They correspond to KOI-1958.01
  and KOI-2115.01 \citep[also Kepler-66\,b and
    Kepler-67\,b;][]{Meibom:2013}, which are the first two transiting
  planets discovered in an open cluster (NGC~6811). The
  \blender\ validations for these candidates adopted much more
  conservative assumptions than usual about the planet prior because
  of the poorly known planet occurrence rate in clusters, and this
  resulted in significantly higher FPP values than would have been
  obtained if the host stars had been in the field.} Under the
conditions of this comparison (i.e., excluding HTP and BP scenarios)
all 34 candidates are validated with \blender\ at the 99.73\%
(3$\sigma$) confidence level or higher ($\log {\rm FPP} < -2.57$;
dashed line), in the sense of indicating the probable presence of a
planet somewhere in the system (not necessarily orbiting the
target). The success rate for \vespa\ (at the 99\% confidence level)
for the same sample is about 70\%.  On the other hand, by being
computationally simpler \vespa\ is designed to allow the quick and
automatic application to large numbers of \kepler\ candidates,
something that is not practical with \blender.

The statistical validation of small candidate transiting planets in or
near the HZ is especially challenging for any method because the
signals are small and the periods tend to be long, resulting in fewer
transits over the duration of the observations. Both of these
contribute to lower signal-to-noise ratios, which results in a less
well defined shape for the transit with which to rule out blends.
These types of candidates benefit the most from close attention to the
circumstances of each one, and the use of every available piece of
information from follow-up observations that can help to exclude
scenarios that could mimic transits. Still, the possibility always
remains that further scrutiny may reveal a ``validated'' planet to be
a false positive of one kind or another. Several examples of such
``unvalidated'' or disproven transiting planets have appeared in the
recent literature.  Three cases from the {\it K2\/} Mission (K2-78\,b,
K2-82\,b, and K2-92\,b) that had been validated as planets with radii
in the 1.4--2.6\;$R_{\earth}$ range \citep{Crossfield:2016} turned out
to be background eclipsing binaries, as shown by \cite{Cabrera:2017}.
In two of the cases the eclipsing binaries were outside the area
covered by the high-resolution imaging observations and had been
overlooked, or produced a clear signature in the flux centroids that
had not been examined, and in the other case there were visible
secondary eclipses that had been missed.  For three other validated
{\it K2\/} planets (K2-51\,b, K2-67\,b, and K2-76\,b) the transit
signals were found to be due to companions that are low-mass stars
rather than planets \citep{Shporer:2017}. In two of them the host
stars had poorly estimated properties, and in the other the size of
the stellar companion is small enough that it is difficult to
distinguish it from a giant planet.  These examples serve as a stark
reminder not only of the probabilistic nature of the validations, but
also of the critical importance of careful attention to detail in the
use of the stellar properties and follow-up constraints.

\acknowledgements

We thank the anonymous referee for helpful comments on the original
manuscript.  This paper includes data collected by the
\kepler\ spacecraft. Funding for the \kepler\ Mission is provided by
NASA's Science Mission Directorate.  Resources supporting this work
were provided by the NASA High-End Computing (HEC) Program through the
NASA Advanced Supercomputing (NAS) Division at Ames Research Center.
The research has also made use of NASA's Astrophysics Data System
(ADS), and of data products from the Mikulski Archive for Space
Telescopes (MAST).  Some of the data presented herein were obtained at
the W.\ M.\ Keck Observatory, which is operated as a scientific
partnership among the California Institute of Technology, the
University of California, and NASA. We extend special thanks to those
of Hawaiian ancestry on whose sacred mountain of Mauna Kea we are
privileged to be guests.  GT acknowledges partial support for this
work from NASA grant NNX14AB83G (\kepler\ Participating Scientist
Program), and Cooperative Agreement NNX13AB58A with the Smithsonian
Astrophysical Observatory (DWL PI).  This research was enabled in part
by support provided by Calcul Qu\'{e}bec (www.calculquebec.ca) and
Compute Canada (www.computecanada.ca).

\end{document}

%% file: planettable1b.tex
\setlength{\tabcolsep}{5pt}
\begin{deluxetable*}{lcc @{\hskip -2pt}c@{\hskip -4pt}cccc}
\tablewidth{0pc}
\tablecaption{Planet Properties Based on the Transit Models.\label{tab:planetproperties1}}
\tablehead{
\colhead{KOI} &
\colhead{$R_p$} &
\colhead{Period} &
\colhead{$T_0$} &
\colhead{$R_p/R_{\star}$} &
\colhead{$b$} &
\colhead{$\rho_{\star}$ ($e=0$)} &
\colhead{S/N}
\\
\colhead{} &
\colhead{($R_{\earth}$)} &
\colhead{(days)} &
\colhead{(BJD$-$2,454,900)} &
\colhead{} &
\colhead{} &
\colhead{(g cm$^{-3}$)} &
\colhead{}
}
\startdata
KOI-0172.01 & 2.36         & 13.722353      & 70.84153     & 0.02211          & 0.0094         & 0.913        & 97.6 \\
         ~  & +0.30/$-$0.21   & $\pm$0.000016  & $\pm$0.00080 & +0.00015/$-$0.00015 & +0.1846/$-$0.0094 & +0.021/$-$0.049 & ~    \\ [0.5ex]
KOI-0172.02 & 1.79         & 242.4659       & 83.8693      & 0.01669          & 0.28           & 0.913        & 21.4 \\
~           & +0.22/$-$0.18   & $\pm$0.0027    & $\pm$0.0080  & +0.00033/$-$0.00049 & +0.10/$-$0.13     & +0.021/$-$0.049 & ~    \\ [0.5ex]
KOI-0438.01 & 1.73         & 5.9311924      & 107.79717    & 0.03065          & 0.0086         & 4.73         & 76.6 \\
~           & +0.28/$-$0.27   & $\pm$0.0000042 & $\pm$0.00059 & +0.00020/$-$0.00023 & +0.1853/$-$0.0086 & +0.19/$-$0.27   & ~    \\ [0.5ex]
KOI-0438.02 & 1.97         & 52.661560      & 116.5188     & 0.03431          & 0.776          & 4.73         & 31.6 \\
~           & +0.19/$-$0.48   & $\pm$0.000111  & $\pm$0.0018  & +0.00043/$-$0.00064 & +0.012/$-$0.015   & +0.19/$-$0.27   & ~    \\ [0.5ex]
KOI-0463.01 & 1.47         & 18.4776440     & 118.26823    & 0.04902          & 0.506          & 18.8         & 71.4 \\
~           & +0.16/$-$0.22   & $\pm$0.0000111 & $\pm$0.00052 & +0.00059/$-$0.00092 & +0.062/$-$0.216   & +5.9/$-$2.8     & ~    \\ [0.5ex]
KOI-0812.01 & 2.169        & 3.3402178      & 104.97911    & 0.03990          & 0.160          & 4.82         & 69.3 \\
~           & +0.078/$-$0.124 & $\pm$0.0000021 & $\pm$0.00058 & +0.00029/$-$0.00029 & +0.057/$-$0.129   & +0.14/$-$0.23   & ~    \\ [0.5ex]
KOI-0812.02 & 2.024        & 20.060390      & 80.4646      & 0.03763          & 0.358          & 4.82         & 34.7 \\
~           & +0.098/$-$0.100 & $\pm$0.000049  & $\pm$0.0020  & +0.00059/$-$0.00059 & +0.063/$-$0.053   & +0.14/$-$0.23   & ~    \\ [0.5ex]
KOI-0812.03 & 2.007        & 46.18406       & 98.2376      & 0.03725          & 0.0041         & 4.82         & 27.7 \\
~           & +0.098/$-$0.100 & $\pm$0.00027   & $\pm$0.0039  & +0.00067/$-$0.00065 & +0.0804/$-$0.0041 & +0.14/$-$0.23   & ~    \\ [0.5ex]
KOI-0812.04 & 1.201        & 7.825046       & 69.5503      & 0.02239          & 0.390          & 4.82         & 14.8 \\
~           & +0.067/$-$0.067 & $\pm$0.000030  & $\pm$0.0038  & +0.00069/$-$0.00069 & +0.072/$-$0.101   & +0.14/$-$0.23   & ~    \\ [0.5ex]
KOI-0854.01 & 2.76         & 56.05623       & 134.1648     & 0.0503           & 0.018          & 5.55         & 26.5 \\
~           & +0.15/$-$0.22   & $\pm$0.00019   & $\pm$0.0026  & +0.0027/$-$0.0027   & +0.105/$-$0.018   & +0.28/$-$0.24   & ~    \\ [0.5ex]
KOI-2418.01 & 1.34         & 86.82974       & 122.2617     & 0.02708          & 0.0068         & 6.13        & 16.7 \\
~           & +0.09/$-$0.14   & $\pm$0.00080   & $\pm$0.0052  & +0.00085/$-$0.00085 & +0.1353/$-$0.0068 & +0.61/$-$0.53   & ~    \\ [0.5ex]
KOI-2626.01 & 1.60         & 38.09722       & 73.0045      & 0.0387           & 0.011          & 9.90         & 16.2 \\
~           & +0.18/$-$0.18   & $\pm$0.00021   & $\pm$0.0047  & +0.0016/$-$0.0018   & +0.233/$-$0.011   & +0.88/$-$1.34   & ~    \\ [0.5ex]
KOI-2650.01 & 1.336        & 34.98956       & 77.2154      & 0.02259          & 0.047          & 4.83         & 12.8 \\
~           & +0.072/$-$0.072 & $\pm$0.00022   & $\pm$0.0063  & +0.00060/$-$0.00106 & +0.093/$-$0.047   & +0.16/$-$0.36   & ~    \\ [0.5ex]
KOI-2650.02 & 1.186        & 7.054259       & 69.1958      & 0.01980          & 0.688          & 4.83         & 11.7 \\
~           & +0.061/$-$0.087 & $\pm$0.000027  & $\pm$0.0037  & +0.00093/$-$0.00092 & +0.028/$-$0.066   & +0.16/$-$0.36   & ~    \\ [0.5ex]
KOI-3010.01 & 2.15         & 60.86628       & 112.2847     & 0.0351           & 0.423          & 4.63         & 16.4 \\
~           & +0.12/$-$0.17   & $\pm$0.00049   & $\pm$0.0069  & +0.0013/$-$0.0019   & +0.060/$-$0.194   & +0.55/$-$0.21   & ~    \\ [0.5ex]
KOI-3282.01 & 1.93         & 49.27668       & 90.0267      & 0.0350           & 0.514          & 5.67         & 24.2 \\
~           & +0.19/$-$0.20   & $\pm$0.00068   & $\pm$0.0117  & +0.0014/$-$0.0024   & +0.078/$-$0.258   & +1.42/$-$0.72   & ~    \\ [0.5ex]
KOI-3497.01 & 0.80         & 20.359722      & 67.2911      & 0.01895          & 0.637          & 10.2         & 35.5 \\
~           & +0.13/$-$0.13   & $\pm$0.000076  & $\pm$0.0025  & +0.00087/$-$0.00101 & +0.075/$-$0.265   & +6.1/$-$2.1     & ~    \\ [0.5ex]
KOI-4036.01 & 1.790        & 168.8116       & 144.4420     & 0.02291          & 0.584          & 2.90         & 28.3 \\
~           & +0.079/$-$0.119 & $\pm$0.0020    & $\pm$0.0075  & +0.00088/$-$0.00088 & +0.051/$-$0.072   & +0.22/$-$0.22   & ~    \\ [0.5ex]
KOI-4054.01 & 2.08         & 169.1335       & 134.8148     & 0.02316          & 0.293          & 2.24         & 27.4 \\
~           & +0.11/$-$0.14   & $\pm$0.0018    & $\pm$0.0096  & +0.00076/$-$0.00076 & +0.075/$-$0.208   & +0.16/$-$0.21   & ~    \\ [0.5ex]
KOI-4356.01 & 2.91         & 174.5085       & 134.482      & 0.0381           & 0.46           & 3.03         & 18.2 \\
~           & +0.17/$-$0.25   & $\pm$0.0029    & $\pm$0.014   & +0.0018/$-$0.0028   & +0.13/$-$0.17     & +0.22/$-$0.22   & ~    \\ [0.5ex]
KOI-4450.01 & 2.23         & 196.4356       & 106.704      & 0.0225           & 0.48           & 1.67         & 16.2 \\
~           & +0.31/$-$0.23   & $\pm$0.0073    & $\pm$0.037   & +0.0017/$-$0.0017   & +0.16/$-$0.27     & +0.24/$-$0.36   & ~    \\ [0.5ex]
KOI-4550.01 & 2.17         & 140.2524       & 108.134      & 0.0286           & 0.52           & 3.08         & 8.9  \\
~           & +0.16/$-$0.23   & $\pm$0.0030    & $\pm$0.017   & +0.0019/$-$0.0027   & +0.13/$-$0.17     & +0.19/$-$0.27   & ~    \\ [0.5ex]
KOI-5236.01 & 2.07         & 550.864        & 240.691      & 0.01801          & 0.52           & 1.00         & 21.8 \\
~           & +0.32/$-$0.20   & $\pm$0.014     & $\pm$0.017   & +0.00076/$-$0.00075 & +0.10/$-$0.26     & +0.41/$-$0.12   & ~    \\ [0.5ex]
KOI-5856.01 & 1.60         & 259.365        & 98.600       & 0.0184           & 0.014          & 0.56         & 13.9 \\
~           & +0.32/$-$0.19   & $\pm$0.014     & $\pm$0.042   & +0.0011/$-$0.0011 & +0.272/$-$0.014   & +0.10/$-$0.17   & ~    
\enddata

\tablecomments{Properties of the KOIs based on our transit light curve
  modeling via MCMC. Uncertainties representing credible 68\%
  confidence intervals are given below each parameter (asymmetric
  error bars are given when warranted).  The main model parameters are
  the period, $T_0$, $R_p/R_{\star}$, $b$, and $\rho_{\star}$. The
  photometric model assumes a circular orbit. The planet radius,
  $R_p$, is the product of $R_p/R_{\star}$ and the stellar radius
  $R_{\star}$ from Table\;\ref{tab:physical}. The signal-to-noise
  ratio S/N was computed using Eq.\,5 of \cite{Rowe:2015}.}

\end{deluxetable*}
\setlength{\tabcolsep}{4pt}

%% file: planettable2b.tex
\setlength{\tabcolsep}{4pt}
\begin{deluxetable*}{lcccc cccc}
\tiny
\tablewidth{0pc}
\tablecaption{Derived Planet Properties Based on our Transit Models.\label{tab:planetproperties2}}
\tablehead{
\colhead{KOI} &
\colhead{$a/R_{\star}$} &
\colhead{Depth} &
\colhead{$\delta_{\rm tot}$} &
\colhead{$\delta_{12}$} &
\colhead{$a$} &
\colhead{$i$} &
\colhead{$S_{\rm eff}$} &
\colhead{$T_{\rm eq}$}
\\
\colhead{} &
\colhead{} &
\colhead{(ppm)} &
\colhead{(hours)} &
\colhead{(hours)} &
\colhead{(au)} &
\colhead{(deg)} &
\colhead{($S_{\earth}$)} &
\colhead{(K)}
}
\startdata
KOI-0172.01 & 20.84      & 591.8     & 5.126      & 0.1109       & 0.1091           & 89.975         & 76          & 867      \\        
         ~  & +0.12/$-$0.50 & $\pm$6.0  & $\pm$0.024  & $\pm$0.0052  & +0.0015/$-$0.0021   & +0.0015/$-$0.459 & +22/$-$16       & +71/$-$90  \\ [0.5ex]
KOI-0172.02 & 141.39     & 333      & 13.10       & 0.2216       & 0.740            & 89.901         & 1.64         & 325      \\        
~           & +0.84/$-$3.37 & $\pm$14  & $\pm$0.25   & $\pm$0.0145  & +0.011/$-$0.013     & +0.011/$-$0.052   & +0.48/$-$0.34   & +35/$-$27  \\ [0.5ex]
KOI-0438.01 & 20.69      & 1035.1    & 2.250      & 0.0672       & 0.0526           & 89.88          & 19.0         & 614      \\        
~           & +0.21/$-$0.49 & $\pm$11.6 & $\pm$0.018  & $\pm$0.0026  & +0.0022/$-$0.0031   & +0.0022/$-$0.37  & +8.6/$-$11.3    & +73/$-$107 \\ [0.5ex]
KOI-0438.02 & 88.72      & 1088     & 3.136       & 0.247        & 0.2254           & 89.495         & 1.03         & 286      \\        
~           & +0.90/$-$2.09 & $\pm$27  & $\pm$0.045  & $\pm$0.013   & +0.0092/$-$0.0139   & +0.0092/$-$0.032  & +0.47/$-$0.61   & +46/$-$41  \\ [0.5ex]
KOI-0463.01 & 69.1       & 2779     & 1.879       & 0.0864       & 0.0899           & 89.606         & 1.18         & 298      \\        
~           & +7.9/$-$2.3   & $\pm$36  & $\pm$0.022  & $\pm$0.0151  & +0.0031/$-$0.0062   & +0.0031/$-$0.094  & +0.22/$-$0.28   & +22/$-$32  \\ [0.5ex]
KOI-0812.01 & 14.15      & 1844     & 1.8672      & 0.07176      & 0.03482          & 89.47          & 44.0         & 731      \\        
~           & +0.13/$-$0.23 & $\pm$25  & $\pm$0.0122 & $\pm$0.00201 & +0.00042/$-$0.00056 & +0.00042/$-$0.33  & +6.5/$-$5.2     & +67/$-$52  \\ [0.5ex]
KOI-0812.02 & 46.76      & 1620     & 3.197       & 0.1334       & 0.1150           & 89.576         & 4.03         & 401      \\        
~           & +0.42/$-$0.76 & $\pm$46  & $\pm$0.054  & $\pm$0.0058  & +0.0014/$-$0.0018   & +0.0014/$-$0.066  & +0.60/$-$0.48   & +38/$-$27  \\ [0.5ex]
KOI-0812.03 & 81.52      & 1619     & 4.484       & 0.1630       & 0.2012           & 89.9971        & 1.33         & 306      \\        
~           & +0.73/$-$1.33 & $\pm$56  & $\pm$0.060  & $\pm$0.0036  & +0.0022/$-$0.0034   & +0.0022/$-$0.0524 & +0.20/$-$0.16   & +24/$-$25  \\ [0.5ex]
KOI-0812.04 & 24.96      & 570      & 2.289       & 0.0578       & 0.06141          & 89.16          & 14.2         & 555      \\        
~           & +0.22/$-$0.41 & $\pm$33  & $\pm$0.071  & $\pm$0.0037  & +0.00074/$-$0.00099 & +0.00074/$-$0.19  & +2.1/$-$1.7     & +43/$-$46  \\ [0.5ex]
KOI-0854.01 & 97.3       & 1851     & 4.574       & 0.221        & 0.2320           & 89.990         & 0.729        & 265      \\        
~           & +1.9/$-$1.1   & $\pm$62  & $\pm$0.067  & $\pm$0.013   & +0.0037/$-$0.0022   & +0.0037/$-$0.060  & +0.091/$-$0.066 & +21/$-$21  \\ [0.5ex]
KOI-2418.01 & 134.6      & 857      & 4.97        & 0.1335       & 0.3006           & 89.9971        & 0.296        & 212      \\        
~           & +5.2/$-$2.9   & $\pm$51  & $\pm$0.14   & $\pm$0.0060  & +0.0069/$-$0.0091   & +0.0069/$-$0.0533 & +0.073/$-$0.037 & +19/$-$17  \\ [0.5ex]
KOI-2626.01 & 91.0       & 977      & 3.24        & 0.1232       & 0.1654           & 89.9927        & 0.81         & 268      \\        
~           & +3.3/$-$3.5   & $\pm$63  & $\pm$0.12   & $\pm$0.0091  & +0.0042/$-$0.0075   & +0.0042/$-$0.1432 & +0.16/$-$0.13   & +27/$-$20  \\ [0.5ex]
KOI-2650.01 & 67.77      & 593      & 4.018       & 0.0896       & 0.1743           & 89.960         & 2.11         & 341      \\        
~           & +0.68/$-$1.84 & $\pm$44  & $\pm$0.078  & $\pm$0.0038  & +0.0023/$-$0.0023   & +0.0023/$-$0.077  & +0.19/$-$0.25   & +27/$-$27  \\ [0.5ex]
KOI-2650.02 & 23.30      & 423      & 1.765       & 0.0628       & 0.05993          & 88.36          & 17.8         & 576      \\        
~           & +0.23/$-$0.63 & $\pm$33  & $\pm$0.087  & $\pm$0.0060  & +0.00079/$-$0.00078 & +0.00079/$-$0.14  & +1.6/$-$2.1     & +55/$-$36  \\ [0.5ex]
KOI-3010.01 & 96.9       & 882      & 4.53        & 0.175        & 0.2543           & 89.750         & 1.06         & 290      \\
~           & +3.7/$-$1.5   & $\pm$58  & $\pm$0.17   & $\pm$0.016   & +0.0043/$-$0.0038   & +0.0043/$-$0.052  & +0.11/$-$0.15   & +21/$-$24  \\ [0.5ex]
KOI-3282.01 & 89.3       & 1377     & 3.79        & 0.138        & 0.2149           & 89.693         & 1.16         & 298      \\
~           & +7.8/$-$3.2   & $\pm$135 & $\pm$0.21   & $\pm$0.023   & +0.0055/$-$0.0054   & +0.0055/$-$0.089  & +0.29/$-$0.31   & +28/$-$28  \\ [0.5ex]
KOI-3497.01 & 60.1       & 317      & 2.094       & 0.0387       & 0.1097           & 89.45          & 1.60         & 322      \\        
~           & +11.4/$-$2.9  & $\pm$19  & $\pm$0.094  & $\pm$0.0128  & +0.0045/$-$0.0079   & +0.0045/$-$0.14   & +0.44/$-$0.51   & +27/$-$40  \\ [0.5ex]
KOI-4036.01 & 163.2      & 601      & 6.61        & 0.219        & 0.5421           & 89.797         & 0.82         & 269      \\        
~           & +4.2/$-$4.1   & $\pm$38  & $\pm$0.26   & $\pm$0.020   & +0.0069/$-$0.0088   & +0.0069/$-$0.022  & +0.13/$-$0.10   & +25/$-$20  \\ [0.5ex]
KOI-4054.01 & 149.5      & 661      & 8.57        & 0.200        & 0.5715           & 89.889         & 1.42         & 309      \\        
~           & +4.2/$-$4.0   & $\pm$40  & $\pm$0.21   & $\pm$0.015   & +0.0069/$-$0.0103   & +0.0069/$-$0.036  & +0.19/$-$0.23   & +29/$-$23  \\ [0.5ex]
KOI-4356.01 & 169.1      & 1759     & 7.64        & 0.320        & 0.5460           & 89.847         & 0.715        & 260      \\        
~           & +4.5/$-$3.7   & $\pm$173 & $\pm$0.44   & $\pm$0.043   & +0.0112/$-$0.0048   & +0.0112/$-$0.050  & +0.113/$-$0.089 & +24/$-$19  \\ [0.5ex]
KOI-4450.01 & 151.8      & 605      & 9.40        & 0.225        & 0.6421           & 89.814         & 1.56         & 324      \\        
~           & +6.9/$-$11.1  & $\pm$80  & $\pm$0.81   & $\pm$0.046   & +0.0108/$-$0.0089   & +0.0108/$-$0.065  & +0.36/$-$0.31   & +28/$-$32  \\ [0.5ex]
KOI-4550.01 & 146.7      & 665      & 6.60        & 0.236        & 0.4706           & 89.788         & 0.98         & 284      \\        
~           & +4.2/$-$3.2   & $\pm$80  & $\pm$0.52   & $\pm$0.038   & +0.0098/$-$0.0041   & +0.0098/$-$0.053  & +0.18/$-$0.10   & +25/$-$21  \\ [0.5ex]
KOI-5236.01 & 276.4      & 373      & 14.76       & 0.273        & 1.297            & 89.889         & 0.77         & 273      \\        
~           & +9.5/$-$31.2  & $\pm$27  & $\pm$0.51   & $\pm$0.056   & +0.014/$-$0.020     & +0.014/$-$0.032   & +0.21/$-$0.16   & +23/$-$27  \\ [0.5ex]
KOI-5856.01 & 127.3      & 395      & 15.73       & 0.291        & 0.745            & 89.9954         & 1.17          & 304      \\        
~           & +4.9/$-$16.1  & $\pm$45  & $\pm$1.08   & $\pm$0.046   & +0.021/$-$0.020     & +0.0021/$-$0.0844   & +0.58/$-$0.28     & +39/$-$31  
\enddata

\tablecomments{Additional derived properties of our KOIs based on our
  transit light curve modeling via MCMC (see
  Table\;\ref{tab:planetproperties1}).  Uncertainties representing
  credible 68\% confidence intervals are given below each parameter
  (asymmetric error bars are given when warranted).  The values of
  $a/R_{\star}$ and the incident flux ($S_{\rm eff}$) were calculated
  using Eq.\,2 and Eq.\,4 from \cite{Rowe:2015}.  $\delta_{\rm tot}$
  is the total depth of the transit model evaluated at time $T_0$, and
  $\delta_{12}$ is the duration of ingress (first to second
  contact). The values of $\delta_{\rm tot}$ and $\delta_{12}$ were
  derived from Eq.\,3 and Eq.\,2 of \cite{Seager:2003}.}

\end{deluxetable*}
\setlength{\tabcolsep}{6pt}